\documentclass[fleqn,usenatbib]{mnras}
\usepackage[english]{babel}
\usepackage[T1]{fontenc}
\usepackage{ae,aecompl}
\usepackage{fancyhdr}
\usepackage{amsfonts}
\usepackage{amsmath}
\usepackage{amssymb}
\usepackage{multicol}
\usepackage{layout}
\usepackage{graphicx}
\usepackage{multirow}
\usepackage{color}
\usepackage{multirow}
\usepackage{times}
\usepackage{natbib}
\def\be{\begin{equation}}
\def\ee{\end{equation}}

\usepackage[utf8]{inputenc}
\usepackage{lmodern}

\definecolor{valecol}{rgb}{0,0.5, 1.}
\newcommand{\vale}[1]{\textcolor{valecol}{#1}}

\newif\ifAMStwofonts
\AMStwofontstrue

\graphicspath{{./fig/}}

\title[Cosmological constrains on minimally and non-minimally coupled scalar field models]{Cosmological constrains on minimally and non-minimally coupled scalar field models}

\author[Davari, Marra and Malekjani]{
	Zahra Davari$^{1}$, Valerio Marra$^2$ and Mohammad Malekjani \thanks{malekjani@basu.ac.ir}$^{3}$ \\ 
	$^1$ Department of Physics,Sharif University of Technology, P.O.Box 11365-9161, Tehran,
	Iran\\
	$^2$ N\'ucleo Cosmo-ufes \& Departamento de F\'isica, Universidade Federal do Esp\'irito Santo, 29075-910, Vit\'oria, ES, Brazil\\
	$^3$ Department of Physics, Bu-Ali Sina University, Hamedan 65178, Iran}
\date{Accepted ?, Received ?; in original form \today}

\pagerange{\pageref{firstpage}--\pageref{lastpage}} \pubyear{2019}

\begin{document}
	\label{firstpage}
	
	\maketitle	
	\begin{abstract}	
We study the minimally and non-minimally coupled scalar field models as possible alternatives for dark energy, the mysterious energy component that is driving the accelerated expansion of the universe.
After discussing the dynamics at both the background and perturbation level, we confront the two models with the latest cosmological data.
After obtaining updated constraints on their parameters we perform model selection using the basic information criteria.
We found that the $\Lambda$CDM model is strongly favored when the local determination of the Hubble constant is not considered and that this statement is weakened once local $H_0$ is included in the analysis. We calculate the parameter combination $S_8=\sigma_8\sqrt{\Omega_{m}/0.3}$ and show the decrement of the tension with respect to the Planck results in the case of minimally and non-minimally coupled scalar field models. Finally, for the coupling constant between DE and gravity, we obtain  the constraint $\xi\simeq -0.06^{+0.19}_{-0.19}$, approaching the one from solar system tests $|\xi| \lesssim 10^{-2}$ and comparable to the conformal value $\xi=1/6$ at  $1\sigma$ uncertainty.
	\end{abstract}
	
	
	
	\section{Introduction} \label{sec:intro}
	During the last decades, cosmological observations have been strongly indicating that the universe is undergoing a process of accelerated expansion.
	This conclusion is supported by both direct geometric probes (such as type Ia supernovae (SnIa) \citep{Betoule:2014frx,Scolnic:2017caz}, the angular location of the first peak of the CMB power spectrum \citep{Hinshaw:2012aka,Aghanim:2018eyx} and baryon acoustic oscillations of the matter power spectrum \citep{Kazin:2014qga,Alam:2016hwk}) and dynamical probes of the growth rate of matter perturbations (such as power spectrum of Ly-$\alpha$ forest at various redshift slices \citep{McDonald:2004xn,Slosar:2013fi}, weak lensing surveys \citep{Kohlinger:2017sxk,Abbott:2017wau} and redshift distortions observed through the anisotropic pattern of galactic redshifts \citep{Blake:2012pj,Pezzotta:2016gbo,Wang:2017wia,Zhao:2018jxv}).
	Understanding the origin of cosmic acceleration is a significant opportunity for theorists, as this cosmic expansion scenario implies that the total energy density today is dominated by an additional cosmic fluid with a positive energy density and a sufficiently negative pressure, usually dubbed the dark energy (DE) \citep{Peebles:1998qn}.
	The nature of DE is one of the greatest unsolved problems in cosmology meanwhile one of the most intriguing questions, and it seems that new physics beyond the standard model of particle physics is required to explain its properties.
	
	One can classify logical possibilities for the nature of DE as follows \citep{Gannouji:2006jm}:
\begin{itemize}
	\item [1-]
	The simplest hypothesis for the DE is to associate it with a positive cosmological constant in Einstein's equations, with $\rho_{DE}=-P_{DE}=\Lambda/8\pi G=$ const. 
	Although successful in fitting available data, the cosmological constant model ($\Lambda$CDM) has a number of theoretical shortcomings such as the fine tuning problem (i.e., the fact that the value of this cosmological constant inferred from observations is extremely small 
	 compared with the energy scales	of high energy physics (Planck, grand unified theory, strong and even electroweak scales)) and the coincidence problem (why this kind of exotic matter starts to dominate today)~\citep{Weinberg:1988cp,Carroll:2000fy, Padmanabhan2003, Copeland2006}.
		In addition to these conceptual and theoretical problems,
		 a tension between CMB observations \citep{Aghanim:2018eyx} and some independent observations at intermediate cosmological scales $(z \leq 0.6)$ has recently been revealed.
		Such tensions include estimates of the Hubble parameter \citep{Riess:2019cxk}, the amplitude of the power spectrum on the scale of $8h^{-1} $ Mpc $(\sigma_8 )$ and the matter density parameter $\Omega_{m}$ \citep{Ade:2015fva,Abbott:2017wau}.
		Therefore, despite the simplicity of this model, it may not be the best description for cosmic acceleration.
		
		\item[2-]	 
		Physical DE: DE is a new cosmological component with negative equation of state, often described through the dynamics of a self-interacting, minimally coupled, scalar field that slowly rolls down its almost flat self-interaction  potential.
		 It is basically a time-dependent cosmological constant. 
		 This family of models (the so-called $\phi$CDM model) has the potential to avoid the fine tuning problem because of the existence of tracking solutions and allows the initial conditions for the scalar field to vary by about 150 orders of magnitude \citep{Brax:1999yv, Zlatev:1998tr}, having a more natural explanation for the observed low energy scale of DE \citep{Ratra:1987rm, Wang:1998gt, Riazuelo:2001mg}. These scalar field models can be classified via their barotropic equation of state parameter ($w=P_{\phi}/\rho_{\phi}$). The models with $-1 < w <-1/3$ are referred to as quintessence models, while the models with $w < -1$ are referred to as phantom models \citep{Davari:2018wnx}. The quintessence models can be classified dynamically in two broad classes: tracking quintessence, in which the evolution of the scalar field is slow, and thawing quintessence, in which the evolution is fast compared to the Hubble expansion. 
		In tracking models the scalar field exhibits tracking solutions in which the energy density of the scalar field scales as the dominant component at the time; therefore the DE is subdominant but closely tracks first  radiation and then matter for most of the cosmic evolution. At some point in the matter domination epoch the scalar field becomes dominant, which results in an effective negative pressure and accelerated expansion.
		These scalar fields might be coupled to the other components of the Universe (matter, for example).
		Within these models one does not expect to observer any anisotropic stress.

		\item[3-]
		Geometric DE: the Einstein general relativity (GR) equations are not the correct
		ones for gravity, but we write them in the Einsteinian form by convention, putting all arising additional terms into the r.h.s.~of the equations and calling them the effective energy-momentum tensor of DE.
An example is the so-called $f(R)$ gravity where the function of the Ricci scalar appearing in the gravitational sector of the Lagrangian is modified.
		See, e.g., \citet{Chiba:1999jz, Capozziello:2003gx,Hu:2007nk}.
Another example is when a coupling between  the scalar field $\Phi$ and the Ricci scalar is allowed, the so-called non-minimally coupled quintessence models or scalar-tensor (ST) theories.
		For a historical review 
		see~\citet{Brans1961,Perrotta:1999am,Pettorino:2008ez}.
		These models typically predict a nonzero anisotropic stress.

	\end{itemize}
	
	Of course, this classification is not absolute. In this work, we intend to study minimally and non-minimally coupled scalar field models, the so-called ``scalar-tensor cosmology.''
	In particular, we wish to obtain the observational constraints on these quintessence models and also to assess if these models could solve the tensions mentioned above.
	
	This paper is organized as follows: In section \ref{sec1}, we first introduce the ST theories of gravity, and then describe the evolution of the background cosmologies in minimally and non-minimally coupled scalar field models. 
	In section \ref{sec2} we present the basic equations governing the evolution of DM at linear perturbation level.
	In section \ref{sec3}, by using the latest cosmological data at background and perturbation level, we perform a statistical analysis in order to constraint the
	free parameters of the models and compare them to each other and to the $\Lambda$CDM model. Finally, we conclude in section \ref{sec5}.
	\section{Background Evolution}\label{sec1}
	In this section, we want to provide a brief review of ST
	models. Since our Universe is homogeneous and isotropic to a high degree on large scales, a Friedmann-Lemaître-Robertson-Walker (FLRW) metric is a very good approximation for the universe we live in:
	\begin{align}\label{frw}
	ds^2 &= -dt^2 + a^2(t)\delta_{ij}dx^idx^j \\
	&= -dt^2 + a^2(t) \left[{dr^2}+r^2 \left(d\theta^2+\sin^2{\theta}~d\phi^2\right)\right], \nonumber
	\end{align}
	where $a(t)$ denotes the cosmic scale factor.
	
	In order to guarantee the weak equivalence principle the matter Lagrangian $\mathcal{L}_m$ must be independent from the scalar field $\Phi$. The Lagrangian density of the scalar field in the Jordan frame is described by~\citep{Bergmann:1968ve, Wagoner:1970vr}
	\begin{equation}\label{lan}
	\mathcal{L}_{\Phi}=\sqrt{-g} \left [\frac{1}{2}(F(\Phi)R-Z(\Phi)g^{\mu\nu}\partial_{\mu}\Phi\partial_{\nu}\Phi)-U(\Phi) \right],
	\end{equation}
	where $R$ is the Ricci scalar, $F(\Phi)$ and $Z(\Phi)$ are arbitrary dimensionless functions, and $U(\Phi)$ is the scalar field potential. The dynamics of the real scalar field $\Phi$ depends on the dimensionless functions $F(\Phi)$ and $Z(\Phi)$ as well as the potential $U(\Phi)$.
	
	The term $F(\Phi)R$ represents the non-minimally coupling between the scalar field $\Phi$ and gravity. $F(\Phi)$ is a dimensionless function which needs to be positive to ensure that the graviton carries positive energy and $dF/d\Phi < 4 \times 10^{-4}$ according
	to Solar System tests. In the GR limit of $F(\Phi)=1$  there is no direct interaction between the scalar field and gravity. For the non-minimally coupled  scalar field models or the extended quintessence (EQ), following \cite{Sanchez:2010ng}, we consider the following general function $F(\Phi)$:
	\begin{equation}
	F(\Phi)=\beta + (1-\beta)cos^{2}(\sqrt{\lambda}\Phi)  ~~ \lambda >0 \label{Fphi-cos}.
	\end{equation}
This form of $F(\Phi)$ is consistent with solar system tests for $\Phi \sim 0$ \citep{Williams:1995nq}. Moreover, this $F(\Phi)$ is similar to the one reconstructed from SnIa observational data where $F(\Phi)$ never becomes negative and does not lead to instabilities \citep{Perivolaropoulos:2005yv}. For small values of $\Phi$, we can expand Eq.(\ref{Fphi-cos}) around $\Phi=0$ and keep terms up to $\Phi^2$. Therefore, in our numerical analysis, we use the following simple form of $F(\Phi)$:
\begin{equation}
F(\Phi)= 1 + \xi \Phi^2 \label{Fphi}.
\end{equation}
where $\xi = (\beta -1)\lambda$, the coupling constant between DE and gravity, is assumed to be dimensionless. A positive coupling $\xi>0$ indicates a flow of energy from gravity to DE and vice versa for $\xi<0$. Ordinary or minimally coupled quintessence  (OQ) is recovered in the limit ${\xi} \rightarrow 0$. Solar system tests constrain the coupling parameter to $\lesssim 10^{-2}$ (see \citealt{Reasenberg:1979ey,Bertotti2003}). It has been shown that for $\Phi <0.2$, the above two forms of $F(\Phi)$ are practically identical \citep{Sanchez:2010ng}.

\begin{figure*}
	\begin{center}
		\includegraphics[width=8.7cm]{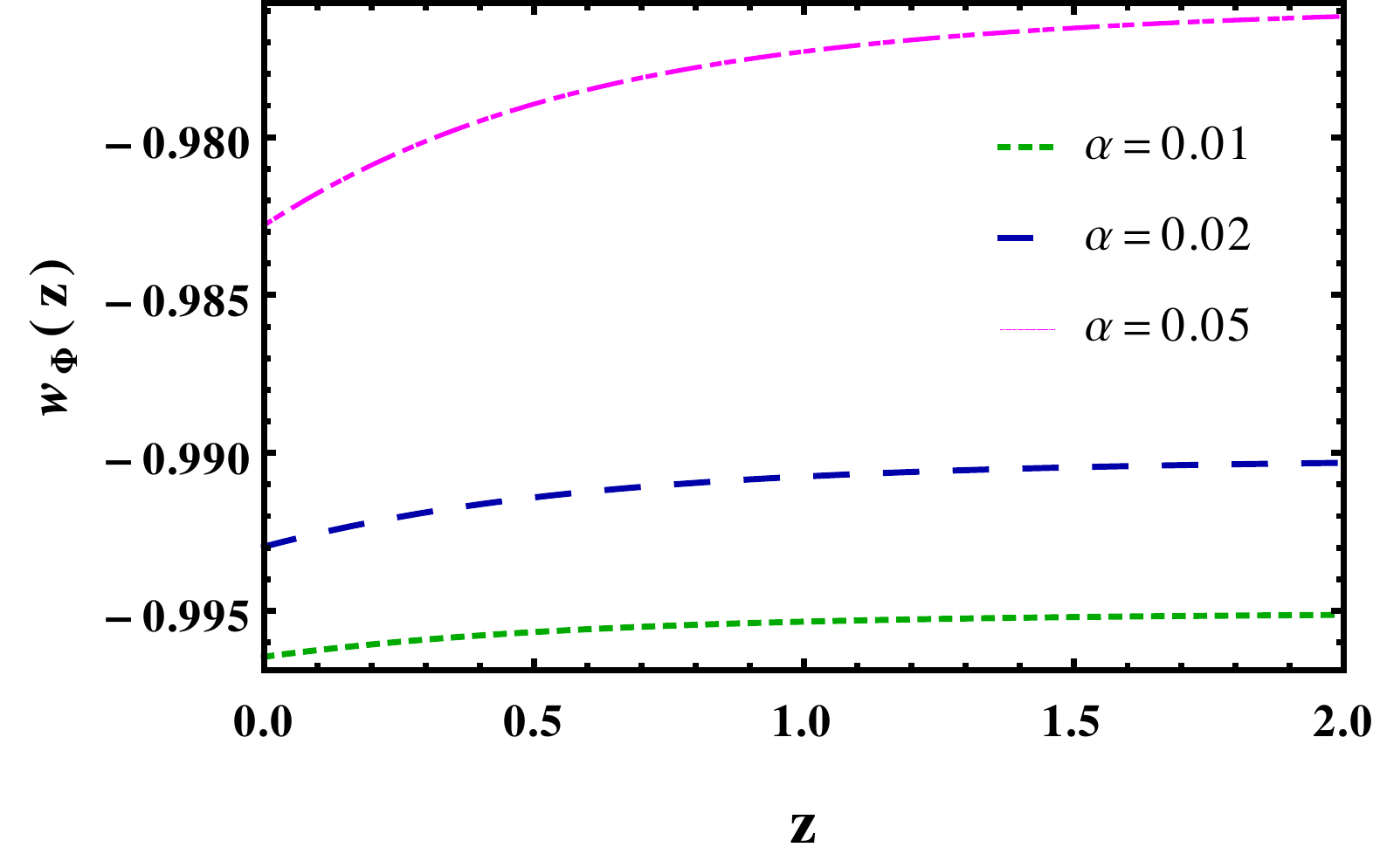}
		\includegraphics[width=8.7cm]{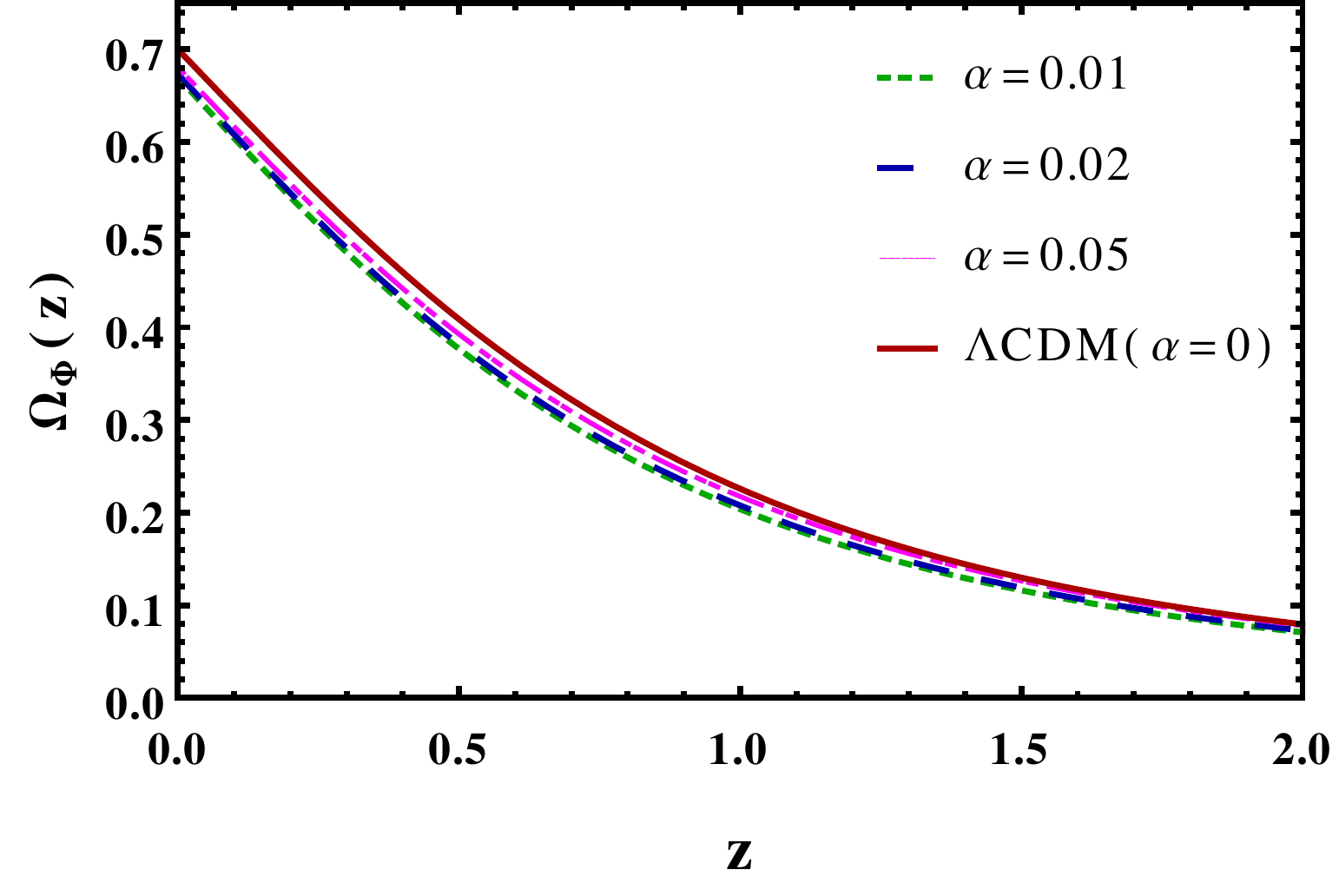}
		\includegraphics[width=8.7cm]{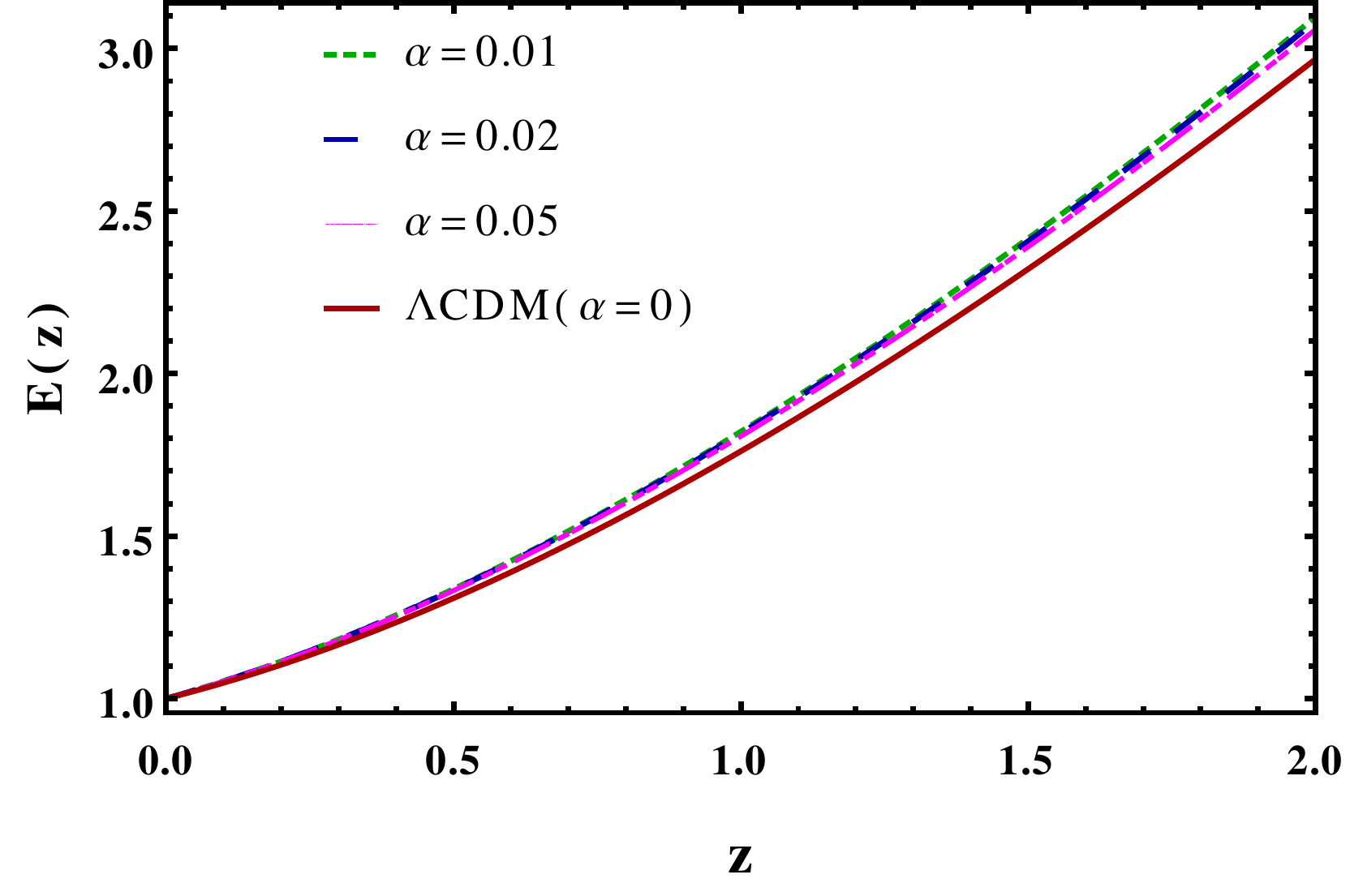}
		\includegraphics[width=8.7cm]{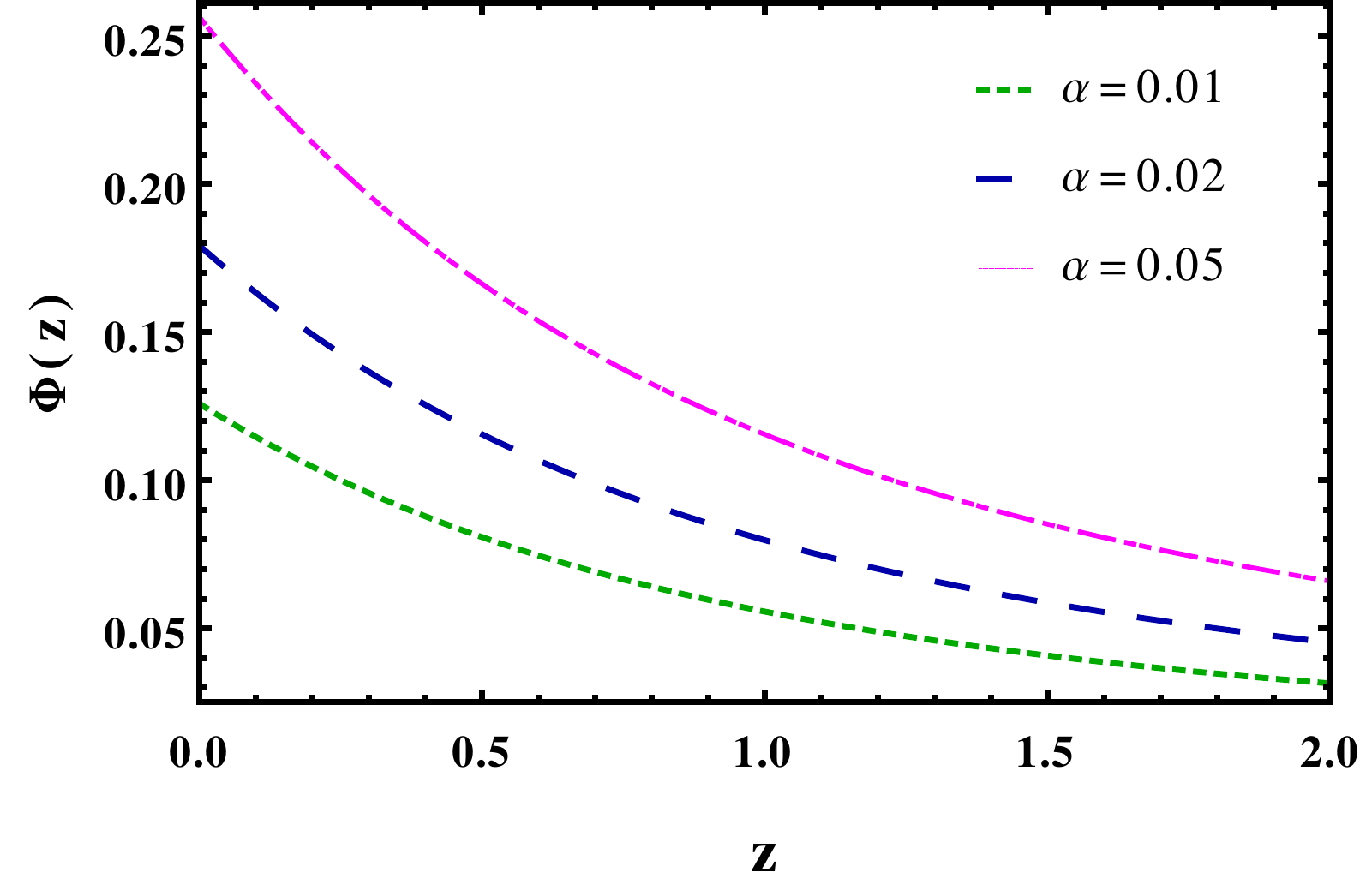}
		\caption{Top left panel: The redshift evolution of the equation of state $w_{\Phi}$.
			Top right panel: The redshift evolution of the energy density parameter of the scalar field ($\Omega_{\Phi}$). Bottom left panel: The redshift evolution of the normalized Hubble parameter $E(z)$. Bottom right panel: The redshift evolution of the scalar field $\Phi$. The  dotted, dashed and dotted dashed curves
			correspond to MSF models with $\alpha=0.01,0.02$ and 0.05, respectively. Note
			that the reference $\Lambda$CDM model is shown by solid red curve.}\label{fig:modelmsf}
	\end{center}
\end{figure*}	

	By a redefinition of the field $\Phi$, the quantity $Z(\Phi)$ can be set to either 1 or $-1$, apart from the exceptional case $Z(\Phi)\equiv 0$  when the scalar-tensor theory  reduces to the higher-derivative gravity theory $R+f(R)$. In this work we will consider the case $Z(\Phi)=1$, and for the potential we adopt the original Peebles and Ratra (PR) form \citep{Peebles1988}
	\begin{equation}\label{potential}
	U(\Phi)=\frac{1}{2}k M_{pl}^2 \Phi^{-\alpha}.
	\end{equation}
	Here $M_{pl} = 1/\sqrt{G} = 1.22 \times 10^{19} $GeV is the Planck mass in natural units. $k$ and $\alpha$ are dimensionless and nonnegative parameters and will be constrained by the cosmological data. Parameter $\alpha$ describes the steepness of the scalar field potential. Larger values of $\alpha$ correspond to faster evolution of the scalar field and vice versa. The case $\alpha=0$ corresponds to the
	time-independent cosmological constant. Type Ia Supernovae data showed that $\alpha < 1$ at the $1\sigma$ level both for minimally and non-minimally coupled quintessence models~\citep{Caresia:2003ze}. Parameter $k$ sets the mass scale $M$ of the scalar particle
	\begin{equation}
	M\sim \left (\frac{k}{2G} \right)^{\frac{1}{\alpha+4}}.
	\end{equation}

	The Klein-Gordon equation for the evolution of the scalar field can be obtained by varying the Lagrangian with respect to the scalar field:
	\begin{equation}
	\partial_{\mu}\left [\frac{\delta \mathcal{L}_{\Phi}}{\delta (\partial_{\mu}\Phi)} \right ]\!-\!\frac{\delta\mathcal{L}_{\Phi}}{\delta\Phi}=0
	\text{\; or \;} \ddot{\Phi}+3H\dot{\Phi}+\frac{dU}{d\Phi}  =\frac{1}{2}\frac{dF}{d\Phi} R ,  \label{eq:stbphi}
	\end{equation}
	where $R=6(\dot{H}+2H^2)$. By taking the variation of the action  with respect to the metric tensor $g_{\mu\nu}$, one obtains Einstein field equations
	\begin{equation}\label{einseq}
	R_{\mu\nu}-\frac{1}{2}Rg_{\mu\nu}=-\frac{T_{\mu\nu}}{F(\Phi)}.
	\end{equation}
	The energy momentum tensor of the scalar field is given by
	\begin{align}
	T_{\mu\nu}^{(\Phi)}&=-\frac{2}{\sqrt{-g}}\frac{\delta\mathcal{L}_{\Phi}}{\delta g^{\mu\nu}}\\ \nonumber
	&=\partial_{\mu}\Phi\partial_{\nu}\Phi-g_{\mu\nu}[-FR+\frac{1}{2}g^{\alpha\beta}\partial_{\alpha}\Phi\partial_{\beta}\Phi+U(\Phi)].
	\end{align}
	Within FLRW cosmology, one can neglect spatial derivatives $\Phi_{,i}$ as compared to time derivatives $\dot{\Phi}$, and the individual components of the homogeneous part of the energy momentum tensor can be written as
	\begin{align}
	\rho_{\Phi}&=-T_0^{0(\Phi)}=\frac{1}{16\pi G}[\frac{1}{2}\dot{\Phi}^{2}+U(\Phi)-6H\dot{F}(\Phi)]  \label{eq:energy_density}, \\
	P_{\Phi}&=T_i^{i(\Phi)}=\frac{1}{16\pi G}[  \frac{1}{2}\dot{\Phi}^{2}\!\!-\!U(\Phi)\!+\!2(\ddot{F}(\Phi)\!+\!2H\dot{F}(\Phi)]. \label{eq:pressure}
	\end{align}
	We also set $\rho_{\Phi}=\rho_{MC}+\rho_{\xi}$ and $P_{\Phi}=P_{MC}+P_{\xi}$ that 
	correspond to the minimally and non-minimally  coupled parts of the scalar field energy density and pressure. By looking at equations (\ref{eq:energy_density}) and (\ref{eq:pressure}) we can see that the equation of state parameter
	\begin{equation}\label{eqwphi}
	w_{\Phi}=\frac{P_{\Phi}}{\rho_{\Phi}}=\frac{\dot{\Phi}^{2}-2U(\Phi)+4(\ddot{F}(\Phi)+2H\dot{F}(\Phi))}{\dot{\Phi}^{2}+2U(\Phi)-12H\dot{F}(\Phi)},
	\end{equation}
	can become negative if the scalar field evolves slowly in time. The equation of state parameter is always greater than $-1$ and in the freezing limit $\dot{\Phi}^2 \ll U(\Phi)$ one has
	\begin{equation}
	w_{\Phi}\simeq\frac{-2U}{2U}=-1,
	\end{equation}
	so we effectively get a cosmological constant with $ w_{\Phi}\simeq w_{\Lambda}$.
	We require the condition $w_{\Phi} < -1/3$ to realize the late-time cosmic acceleration, which translates into the condition $\dot{\Phi}^2 < U(\Phi)$. We can rewrite equation (\ref{eqwphi}) as:
	\begin{equation}\label{wphi}
	w_{\Phi}=-1+\frac{\alpha n\left[n-2\xi[\alpha(2-n)+2(2+n)]\right]}{3(\alpha+2)\left[n-\xi[(4-n)(\alpha+2)-4n\alpha]\right]}.
	\end{equation}
	The Friedmann equations for the evolution of the background in ST gravity are the following:
	\begin{eqnarray}
	3F(\Phi) H^2  & = &8\pi G( \rho_{\rm m}+\rho_{\Phi})\;, \label{eq:stbh1}\\
	-2F(\Phi) \dot H & = &8\pi G( \rho_{\rm m}+\rho_{\Phi}+p_{\Phi})\;,
	\label{eq:stbh2}
	\end{eqnarray}
	where the Universe was modeled as a mixture of  non-relativistic matter and a scalar field. The density parameter of any component $i$ is defined according to:
	\begin{equation}
	\Omega_i=\frac{8\pi G\rho_i}{3FH^2}.
	\end{equation}
	In the following, we particularize the equations for the minimal and non-minimal scalar field models.
	
		\begin{figure*}
		\begin{center}
			\includegraphics[width=8.7cm]{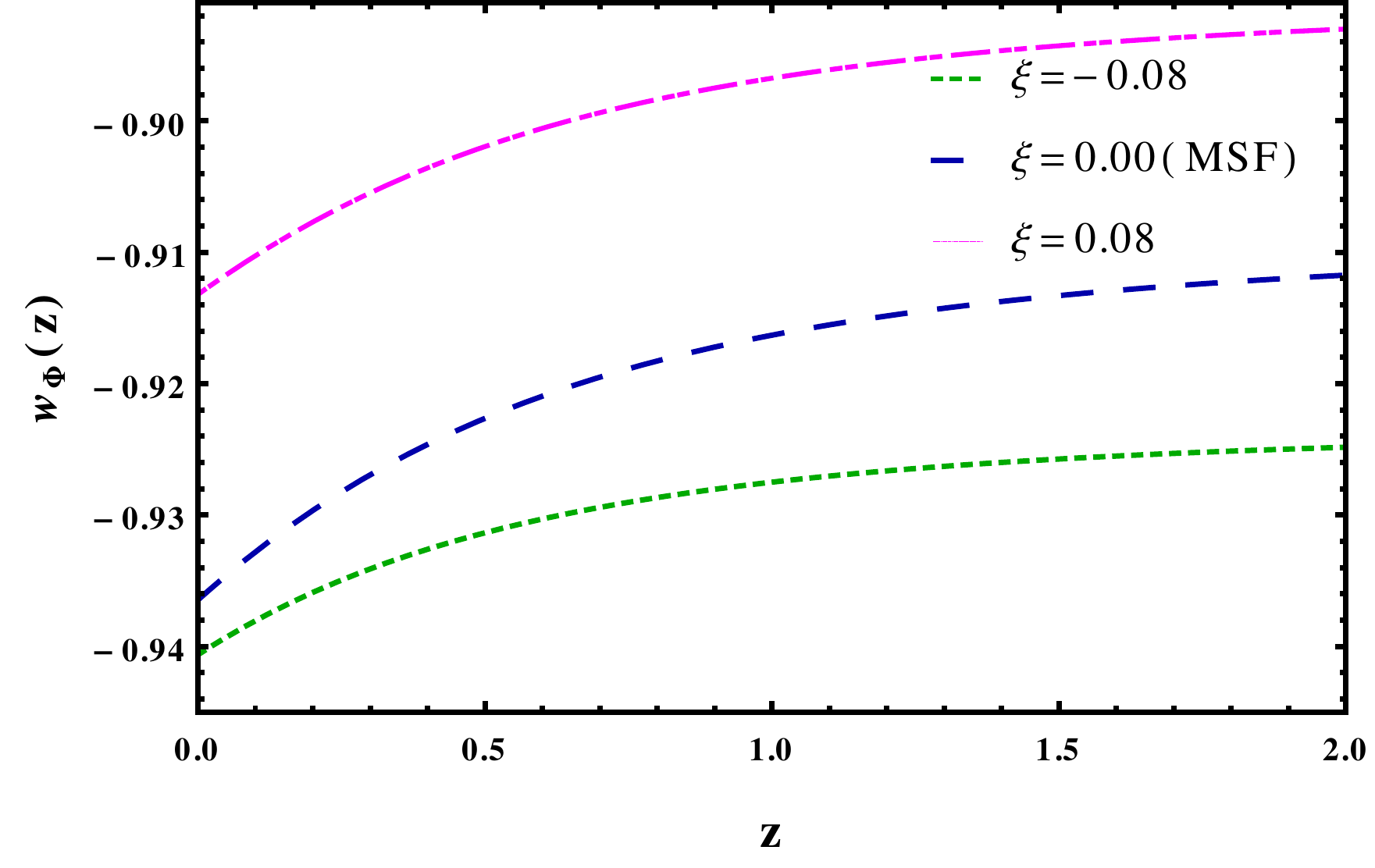}
			\includegraphics[width=8.7cm]{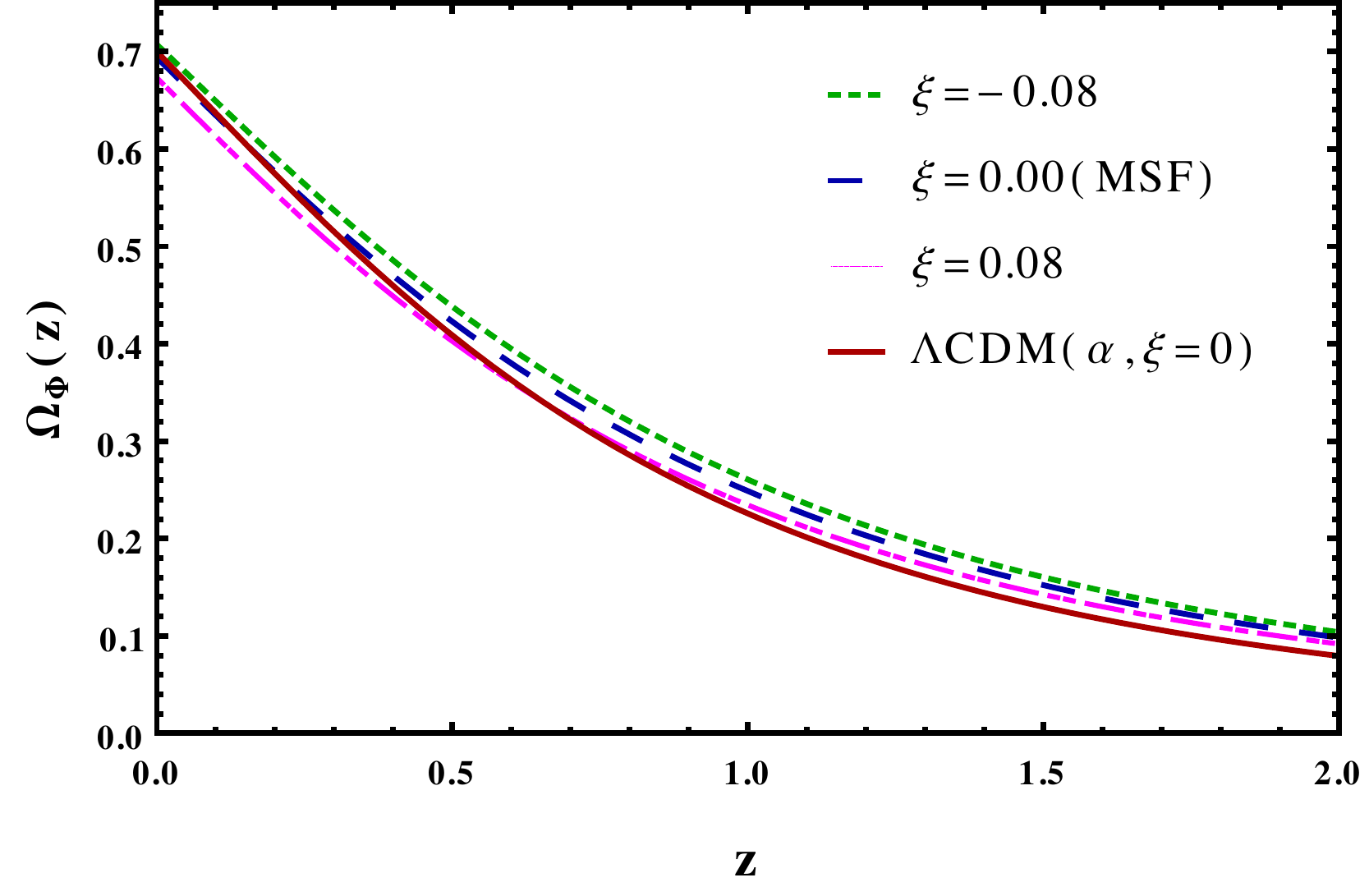}
			\includegraphics[width=8.7cm]{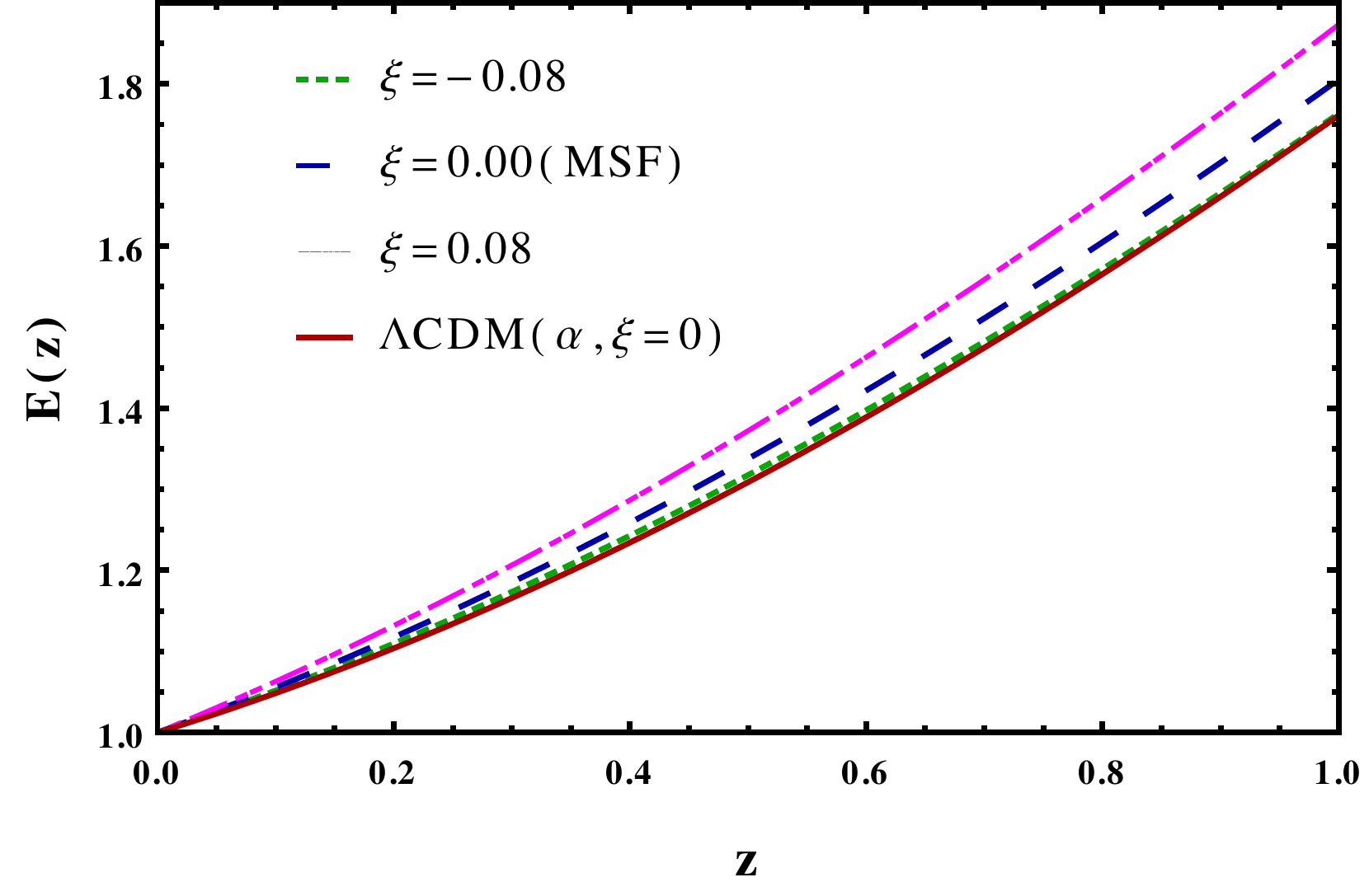}
			\includegraphics[width=8.7cm]{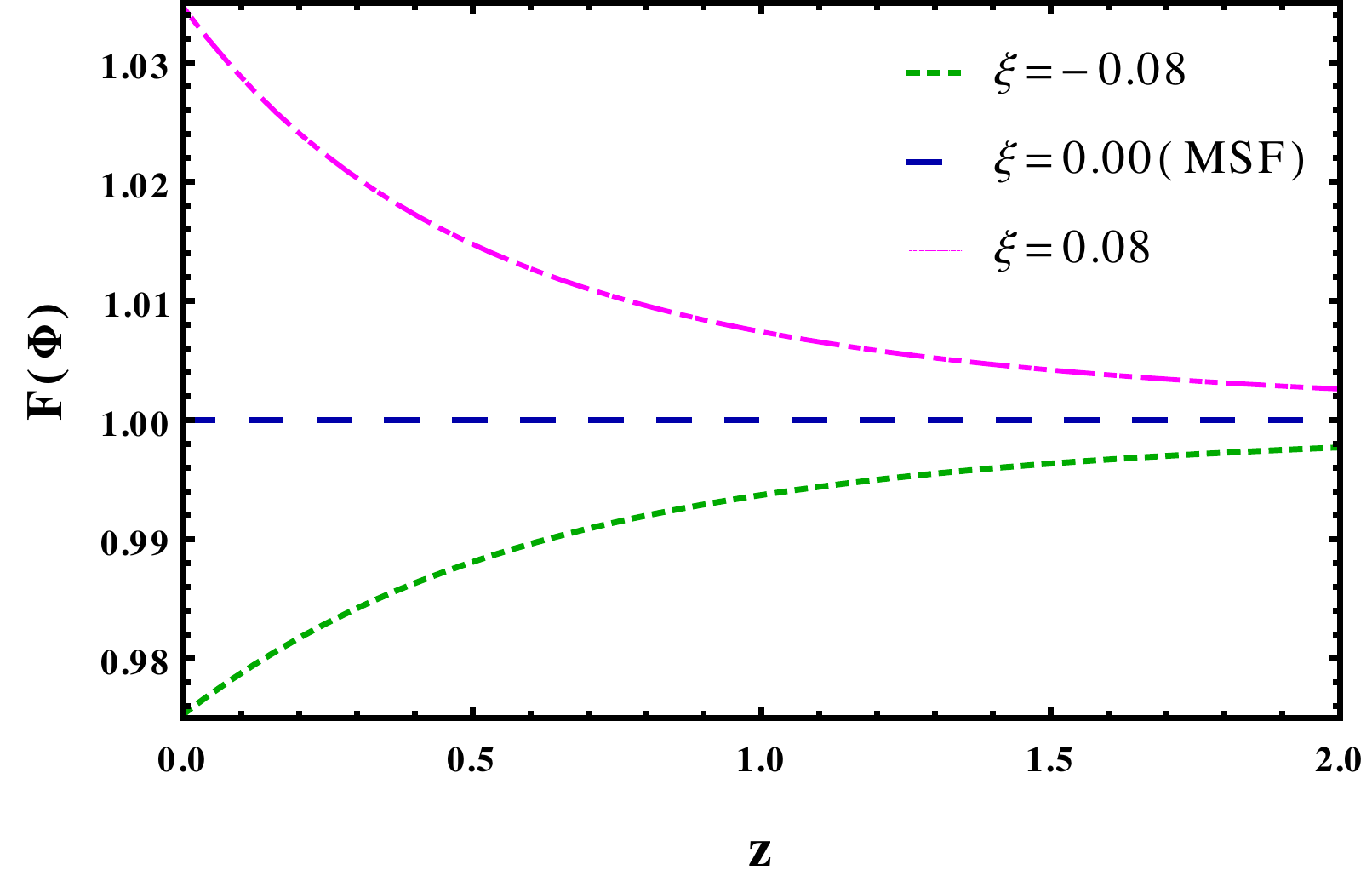}
			\caption{The redshift evolution of $w_{\Phi}$(top left panel),
				$\Omega_{\Phi}$(top right panel), $E(z)$(bottom left panel)  and $F(\Phi)$ (bottom right panel). The  dotted, dashed and dotted dashed curves correspond to MSF models with ${\xi}=-0.08,0$ and 0.08, respectively (we adopted $\alpha=0.2$ for all panels). The red solid curve shows	the concordance $\Lambda$CDM model.}\label{fig:modelnsf}
		\end{center}
	\end{figure*}

	\subsection{Minimal Scalar Field (MSF) or Ordinary Quintessence (OQ)}\label{sec1-1}
In order to solve the Klein-Gordon equation (\ref{eq:stbphi}), we adopt a power-law ansatz to determine the initial conditions for $\Phi$:
\begin{equation} \label{phit}
\phi(t)=A \, t^p \,.
\end{equation}
The continuity equation during matter (MDE) or radiation (RDE) dominated epochs ($i=m,r$) is~\citep{Sola:2016hnq}:
	\begin{equation}
	\frac{ d\rho_{i}}{d a} + \frac{3}{a}(1+w_{i}(a))\rho_{i} = 0,  \label{eq:cont}
	\end{equation}
	from which it is easy to  derive that the evolution of the energy density of dust and radiation are
	$\rho_{\rm m}=\rho_{m0}\, a^{-3}$ and $\rho_{\rm r}=\rho_{r0}\, a^{-4}$, respectively, which can be written collectively as $\rho= \rho_0 \, a^{-n}$ ($n=3$ for MDE and $n=4$ for RDE). Solving Friedmann's equation (\ref{eq:stbh1}) in flat space, we find $H(t) = 2/(nt)$ during the $n$th-epoch ($t$ is the cosmic time). Substituting these relations in the Klein-Gordon equation \eqref{eq:stbphi} with the Peebles-Ratra potential (\ref{potential}) and adopting the ansatz of \eqref{phit} leads to
	\begin{equation}
	\qquad p=\frac{2}{\alpha+2},\qquad A^{\alpha+2}=\frac{\alpha(\alpha+2)^2M_{pl}^2kn}{4(6\alpha+12-n\alpha)}.
	\end{equation}
	We can trade the cosmic time in (\ref{phit}) for the scale factor. Using $H=2/nt$ we obtain $t^2 = 3/(2\pi G n^2 \rho)$ where $\rho(a)=\rho_0 \, a^{-n}=\rho_{c0} \, \Omega_i a^{-n}$,
	$\Omega_i = \Omega_m,\Omega_r$ are the present-day values of the cosmological density parameters and $\rho _{c0}= 3H_0^ 2 /(8\pi G)$  is the current critical energy density. Notice that $\Omega_m = \Omega _{dm} + \Omega _b$ involves both dark matter and baryons. In this way we can determine $\Phi$ as a function of the scale factor during the $n$th-epoch. For example, in the MDE we obtain
	\begin{equation}\label{phia1}
	\Phi(a)=\left[\frac{\alpha(\alpha+2)^2 \bar k}{9\times 10^4 h^2 \Omega_m(\alpha+4)}\right]^{1/(\alpha+2)}a^{3/(\alpha+2)} ,
	\end{equation}
	where the reduced Hubble constant $h$ is defined as usual according to $H_0\equiv 100 \, h \, \zeta$, where $\zeta\equiv 1\text{Km/s/Mpc}=2.133\times 10^{-44}$GeV (in natural units). Finally, for convenience we have introduced in (\ref{phia1}) the dimensionless parameter $\bar{k}$ through $kM_{pl}^2 \equiv\bar{k}\zeta$ .
	
	Using equations (\ref{eq:stbh1}-\ref{eq:stbh2}), and the relations $\bar{H}= H/\zeta$ and $\dot{\Phi}=a H \Phi'(a)$, we can obtain, during the MDE:
	\begin{align}
	\bar{H}^2&=\frac{\bar{k}\Phi^{-\alpha}(a)+1.2\times 10^5 h^2 \Omega_m a^{-3}}{12-a^2\Phi'^2(a)},  \label{hmin} \\
	\bar{H}'&=-\frac{3}{2a\bar{H}}\left[\frac{a^2\bar{H}^2\Phi'^2}{6}+10^4 h^2 \Omega_m a^{-3} \right ],
	\end{align}
	which can be used to solve the following Klein-Gordon equation
	\begin{equation}
	\Phi''+ \left (\frac{\bar{H}'}{\bar{H}}+\frac{4}{a} \right )\Phi'-\frac{\alpha}{2}\frac{k\Phi^{-(\alpha+1)}}{(a\bar{H})^2}=0
	\end{equation}
	with initial conditions from equation (\ref{phia1}), which are expressed in terms of the parameters that enter our analysis. We set these conditions in the MDE at $a_i=0.0001$.
	In figure~\ref{fig:modelmsf}, we show the redshift evolution of $w_{\Phi}(z) $ (top left panel), $\Omega_{\Phi}(z)$ (top right panel) , $E(z)$ (bottom left panel) and  $\Phi(z)$ (bottom right panel) for different values
	of the model parameter $\alpha=0.01$ (dotted line),  $0.02$ (dashed line) and
	0.05 (dot-dashed line). The concordance $\Lambda$CDM cosmology ($\alpha=0$) is also plotted for comparison (see solid line). As expected, the aforementioned cosmological quantities depend on the choice of $\alpha$. We could obtain from equation (\ref{wphi}) for MSF model, $w_{\Phi}=-1+\frac{\alpha n }{3(\alpha+2)} $ and by using  the matter EoS in the $n$th epoch, $w_n=-1+n/3$,  we can rewrite it as $w_{\Phi}=\frac{\alpha w_n-2}{\alpha+2}$. For the matter dominated epoch ($n=3$) we find $w_{\Phi}=\frac{-2}{\alpha+2}$ so that  $w_{\Phi}>-1$ for $\alpha>0$. In figure~\ref{fig:modelmsf} we observe that the EoS parameter remains in the quintessence regime and lies in the interval $-1 \leq w_{\Phi} \leq -1/3$ . Also, for small values of $\alpha$, the EoS parameter tends to $-1$ at the present epoch. The evolution of the energy density parameter of the scalar field shows that for  large values of $z$, $\Omega_{\Phi}$ is larger then the concordance $\Lambda$
	cosmology. Regarding the normalized Hubble parameter $E(z)=H(z)/H(z=0)$, we see that  $E_{MSF}(z) > E_{\Lambda}(z)$.
	It should be noted that in all panels of figure~\ref{fig:modelmsf} we adopted $\{\Omega_{m0}=0.3, h=0.7,\bar k=3.5\times 10^4\}$. Finally, we present the redshift evolution of the scalar field $\Phi$ in the bottom right panel. We see that for all $\alpha$ values, the scalar field increases with time.
		\begin{figure*}
		\begin{center}
			\includegraphics[width=8cm]{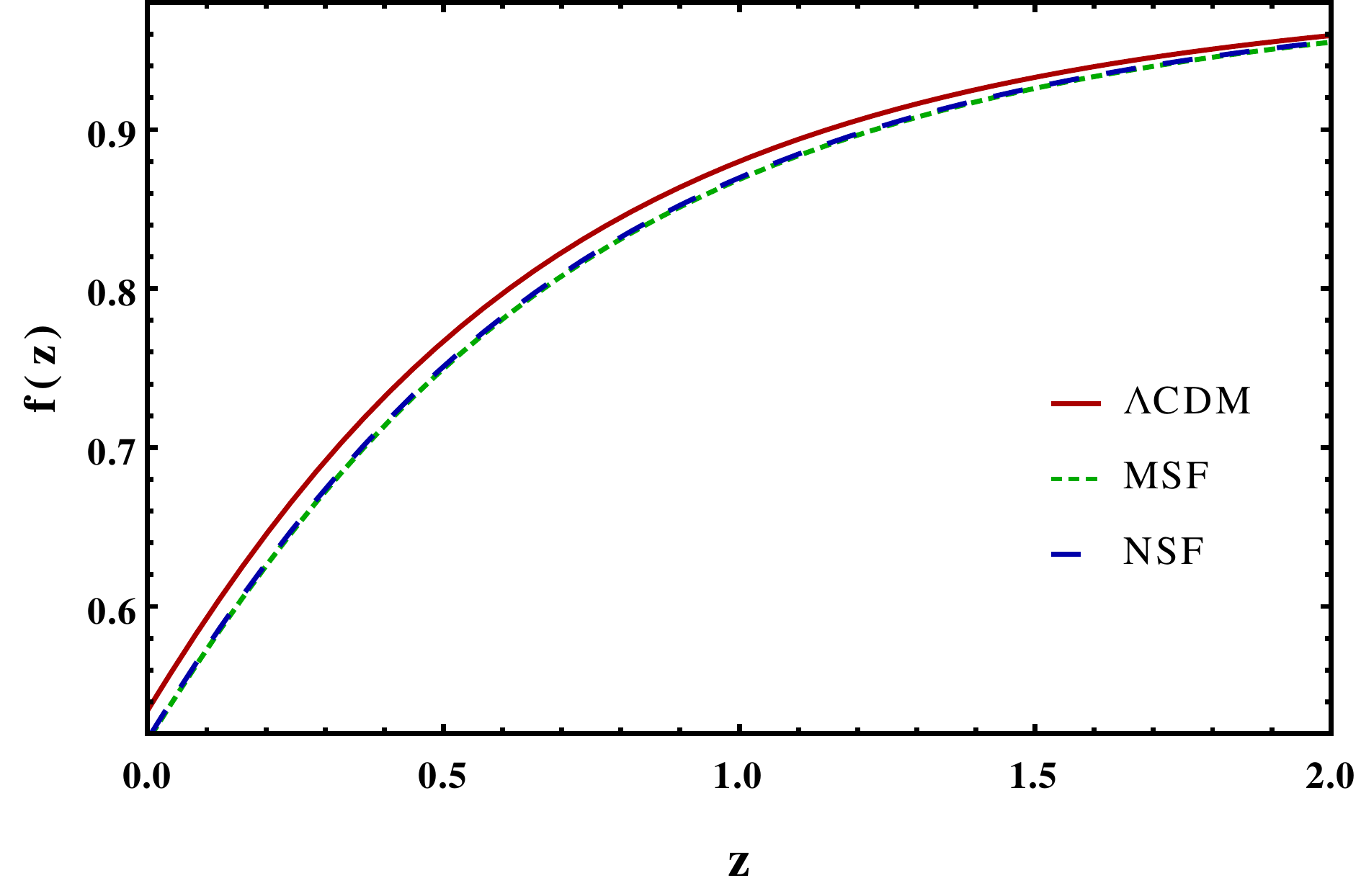}
			\includegraphics[width=8cm]{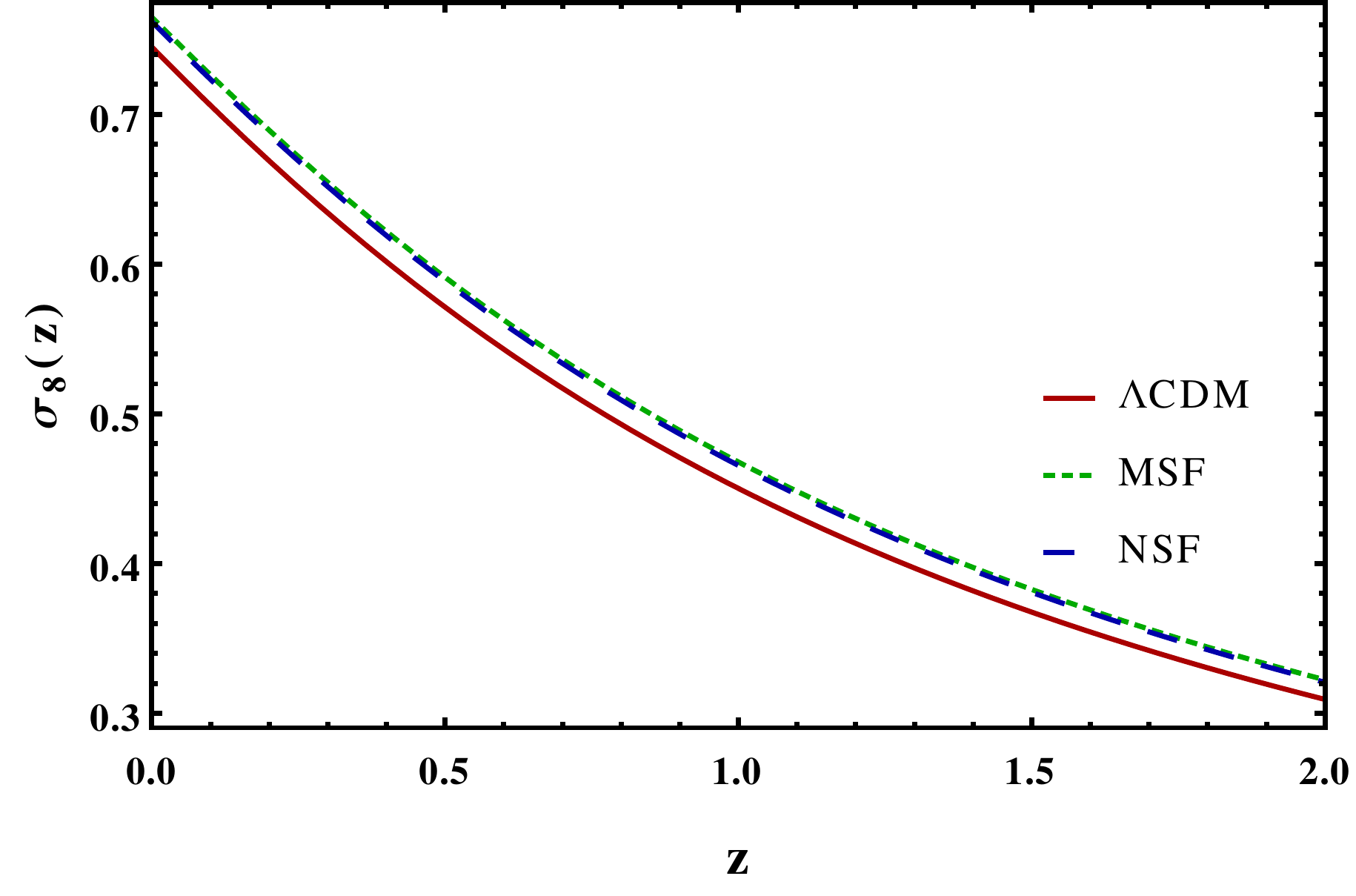}
			\includegraphics[width=8cm]{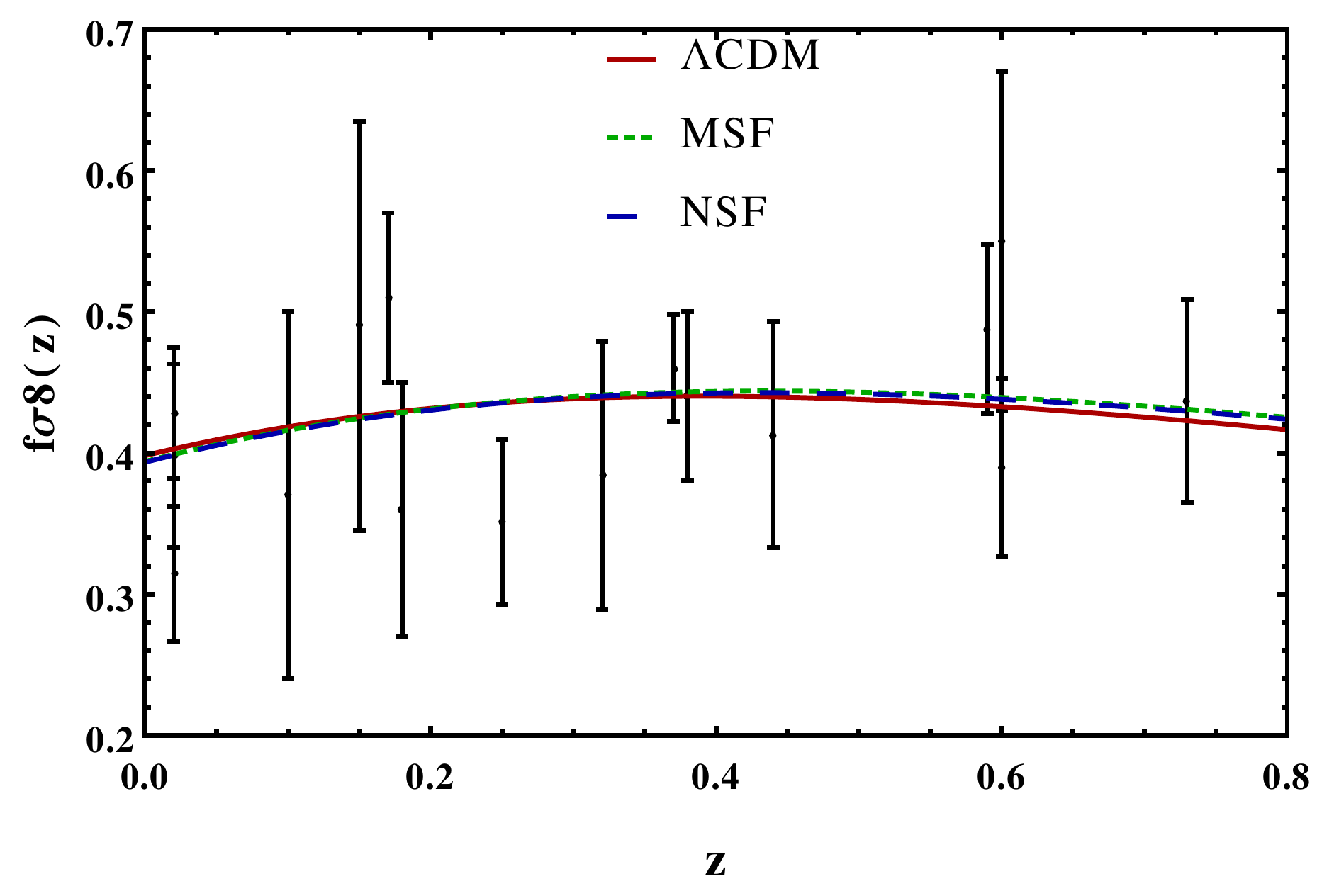}
			\includegraphics[width=8cm]{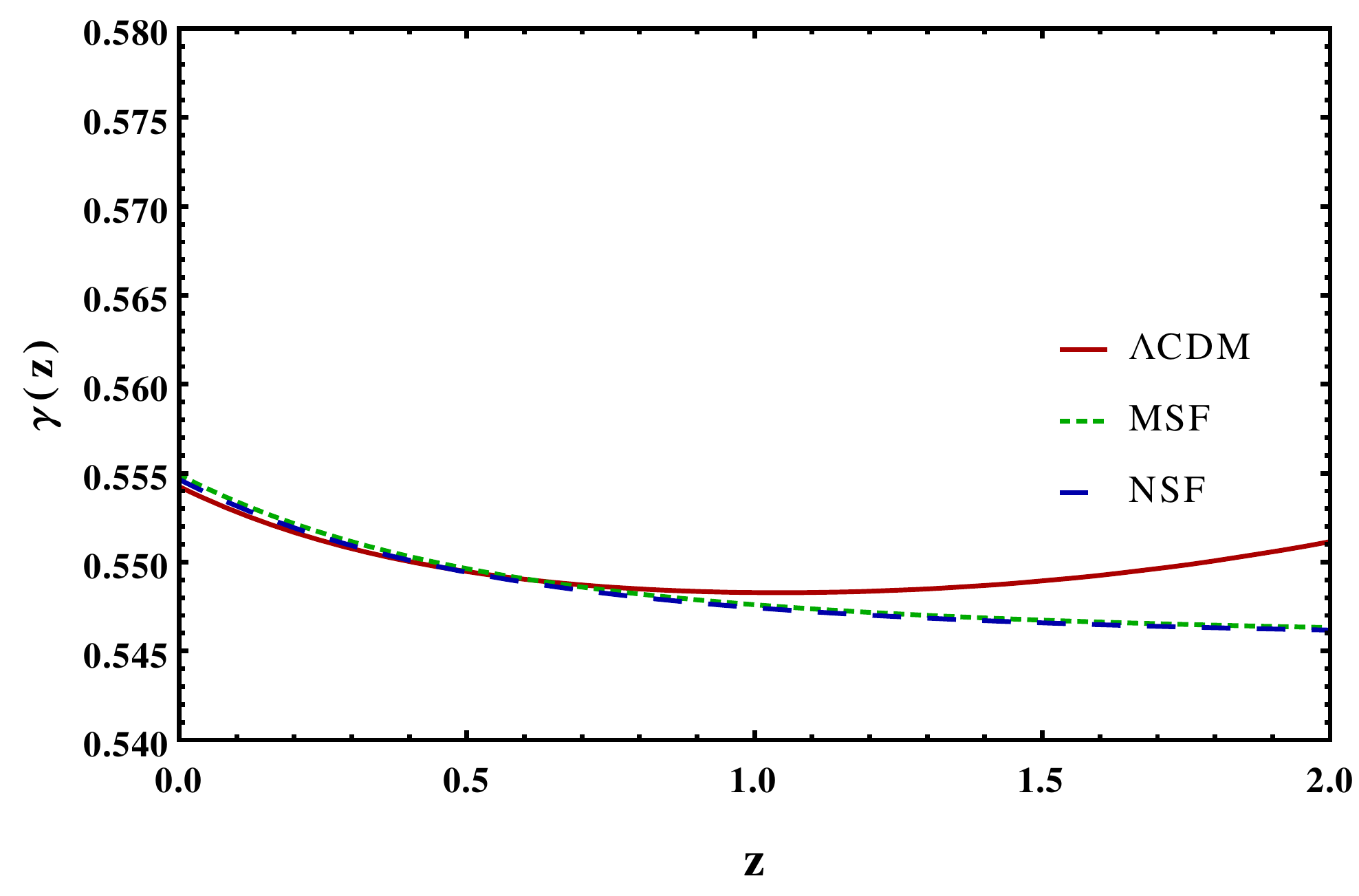}
			\caption{Shown are the redshift evolution of  the growth rate $f(z)$ (top left panel), variance of perturbations $\sigma_8(z)$ (top right panel), of the growth index $\gamma(z)$  (bottom right panel) of matter perturbations. In the bottom left panel we  compare the observed and theoretical evolution of the combination $f(z)\sigma_8(z)$. The red solid, green dotted and  blue dashed curves correspond to $\Lambda $CDM, MSF and NSF models.
				The filled circles are the growth data with their error bars. The cosmological parameters are fixed according to their values given in the lower half of table \ref{tab:res}.
			}\label{fig:grow}
		\end{center} 
	\end{figure*}

	\subsection{Non-Minimal Scalar Field (NSF) or Extended Quintessence (EQ)} \label{sec1-2}
	Here, we study the background evolution for the non-minimally coupled scalar field or extended quintessence model. The coupling between the scalar field and the Ricci scalar is modeled via the term $F(\Phi)R$ in equation (\ref{lan}). Similarly to the minimally coupled scalar field, we  seek power-law solutions to determine the initial conditions, i.e., we adopt the ansatz:
\begin{equation} \label{phitnon}
\phi(t)=B \, t^p \,.
\end{equation}
Substituting these relations in the Klein-Gordon equation and using (\ref{Fphi}-\ref{potential}) one obtains
	\begin{align}
	\vale{p}&=\frac{2}{\alpha+2}, \\
	B^{\alpha+2}&=\frac{\alpha(\alpha+2)^2M_{pl}^2kn^2}{4[n(6\alpha+12-n\alpha)-6{\xi}(4-n)(\alpha+2)^2]} \nonumber \,,
	\end{align}
	which, after trading the cosmic time in (\ref{phitnon}) for the scale factor, becomes during the MDE epoch:
	\begin{equation}\label{phia}
	\Phi(a)\!=\!\left [\frac{\alpha(\alpha+2)^2\bar{k}}{3\times 10^4 h^2 \Omega_m[3(\alpha+4)-2{\xi}(\alpha+2)^2]}\right ]^{\frac{1}{\alpha+2}}a^{\frac{3}{\alpha+2}}.
	\end{equation}
	
	Using equations (\ref{eq:stbh1}-\ref{eq:stbh2}) and the continuity
	equation, we can obtain the evolution of the  Hubble parameter $H(z)$ in the MDE: 
	\begin{align}
	&\bar{H}^2=\frac{\bar{k}\Phi^{-\alpha}(a)+1.2\times 10^5 h^2 \Omega_m a^{-3}}{12F(\Phi(a))\left (1+\frac{a F'(\Phi(a))}{F(\Phi(a))} \right)-a^2\Phi'^2(a)}, \label{hnon} \\
	&\bar{H}'=-\frac{3}{2a\bar{H}F(\Phi(a))(1+\frac{a F'(\Phi(a))}{2F(\Phi(a))})} \label{hpnon} \\
	&\left [\frac{a^2\bar{H}^2\Phi'^2}{6}\!+\!\frac{2{\xi} a^2\bar{H}^2[\Phi'(a)^2+\Phi(a)\Phi''(a)]}{3}\!+\!10^4 h^2 \Omega_m a^{-3} \right] \!.  \nonumber
	\end{align}
	Then, we integrate  the following Klein-Gordon equation
	\begin{equation}
	\Phi''+\left (\frac{\bar{H}'}{\bar{H}}+\frac{4}{a} \right)\Phi'-\frac{\alpha}{2}\frac{k\Phi^{-(\alpha+1)}}{(a\bar{H})^2}=\frac{6{\xi} \Phi}{a} \left (\frac{\bar{H}'}{\bar{H}}+\frac{2}{a} \right)
	\end{equation}
	and the set of equations  (\ref{hnon}-\ref{hpnon})  numerically,
	starting from $a_i =0.0001$ to the present time $a_0 = 1$. 

	
	In figure~\ref{fig:modelnsf} we plot similar panels as in figure~\ref{fig:modelmsf} for  $w_{\Phi}(z) $,
	$\Omega_{\Phi}(z)$, $E(z)$ and  $\Phi(z)$, for different values
	of the model parameter $\xi$. We observe that the EoS parameter at the present epoch tends to $-1$ (the cosmological constant value) more rapidly for negative values of ${\xi}$ than for positive values (top left panel).
	A similar behavior is observed for  $E(z)$ (bottom left panel, behavior closer to the $\Lambda$CDM case).
	For all the cases $\Omega_{\Phi}$  is negligible at high redshifts meaning that for any ${\xi}$ value the model reduces to an EdS Universe at early times.
	In the bottom right panel, we present the redshift evolution of $F(\Phi)$. It shows that  $F\rightarrow1$ for large values of $z$, that is, the non-minimal scalar field model reduces to the minimal scalar field model. 
	It should be noted that in all panels of figure~\ref{fig:modelnsf} we adopted $\{\Omega_{m0}=0.3, h=0.7,\bar{k}=3.5\times 10^4,\alpha=0.2\}$.

	\section{Perturbations In Scalar Tensor Models}\label{sec2}
	A fundamental question in the formation of cosmic structures is how DE and its perturbations could affect the structure formation process. Therefore, we will consider
	linear perturbations of the metric tensor, while keeping non-linearity in the matter and scalar field perturbations. As it is known, metric perturbations can be decomposed into scalar modes, as well as  vector and tensor ones.
	By taking the spatial
	covariant divergence of the $(0-i)$ Einstein equations and the trace of the $(i-j)$ equations, we can single out the scalar
	modes so that, without loss of generality, the line-element can be written, in the  longitudinal gauge, as
	\begin{equation}\label{permetric}
	ds^2 = -(1 + 2\phi)dt^2 + a^2 (1 - 2\psi)\delta _{ij}dx^idx^j,
	\end{equation}
	where $\phi$ and $\psi$ are the linear gravitational potentials which, in the limit of GR and in the absence of
	anisotropic stresses, satisfy $\phi=\psi$. In ST theories of gravity, we can find the connection between the potentials $\phi$ and $\psi$ by using the perturbed metric (\ref{permetric}) in the generalized Einstein equations~(\ref{einseq}):
	\begin{equation} \label{phipsi}
	\phi=\psi-\frac{F{,\Phi}}{F}\delta \Phi,
	\end{equation}
	where $F_{,\Phi}=\frac{dF}{d\Phi}$. For $F=1$ (GR case) this equation reduces to $\phi=\psi$. 
	The linear evolution of non-relativistic matter density perturbations  $\delta_m=\delta\rho_m/\rho_m$ can be obtained via~\citep{Copeland2006,Sanchez:2010ng,Nazari-Pooya:2016bra}
	\begin{equation}
	\ddot{{\delta}}_{\rm m}+2H\dot{{\delta}}_{\rm m}+\frac{k^2}{a^2}
	\left(\psi-\frac{F_{,\Phi}}{F}\delta\Phi\right)-
	3(\ddot{\psi}+2H\dot{\psi})=0 \;,\label{matperteq}
	\end{equation}
	where $k$ is the wave number of the perturbation mode. In the following we consider sub-Hubble ST scales ($\frac{k}{a}\gg H$ and $F_{,\Phi}\geq 1$) so that the perturbation scale and the shifted ST perturbation scale are of the same order. The scalar field perturbation is given by~\citep{Sanchez:2010ng}
	\begin{equation}
	\delta\Phi \simeq (\phi-2\psi) F_{,\Phi}.\; \label{shdphi1}
	\end{equation}
	By substituting equation (\ref{phipsi}), we have
	\begin{equation}
	\delta \Phi\simeq -\psi \frac{F F_{,\Phi}}{F+F_{,\Phi}^2}\;. \label{shdphi2}
	\end{equation}
	Thus, this relation shows that perturbations of non-minimally coupled scalar fields in ST gravity are independent of the scale $k$ on sub-horizon scales. For minimally coupled scalar tensor models, we have $F=1$  then $F_{,\Phi}=0$ and  from equation (\ref{shdphi1}) one has $\delta\Phi=0$: the perturbations of minimally coupled scalar fields on sub-horizon scales are negligible in GR gravity.
	
	It has been shown that  the ratio of non-minimally coupled scalar field perturbations to  matter perturbations at sub-horizon scales is given by~\citep{Sanchez:2010ng}
	\begin{equation}
	\frac{\delta\rho_\Phi}{\delta\rho_{\rm m}}=\frac{\delta_\Phi}{\delta_{\rm m}}\simeq
	-\frac{F_{,\Phi}^2}{3F_{,\Phi}^2+2F}\;. \label{shdrat}
	\end{equation}
	We can then eliminate the term $\frac{k^2}{a^2}\psi$
	from equation (\ref{matperteq}) by using equation (\ref{shdrat}). Also ignoring $\dot{\psi}$ and $\ddot{\psi}$  in equation (\ref{matperteq}), we find that the linear evolution of matter overdensities on sub-horizon scales within ST gravities is governed by
	\begin{eqnarray}
	\ddot{\delta}_{\rm m}+2H\dot{\delta}_{\rm m}-\frac{\rho_{\rm m}\delta_{\rm m}}{2}\frac{1}{F}\left(\frac{2F+4F_{,\Phi}^2}{2F+3F_{,\Phi}^2}\right)=0 \;.\label{per2}
	\end{eqnarray}
	Finally, we can define the effective gravitational constant $G_{\rm eff}$ for the case of homogeneous non-minimally coupled scalar field models as follows
	\begin{equation}
	G_{\rm eff}=\frac{G_{\rm N}}{F}\left(\frac{2F+4F_{,\Phi}^2}{2F+3F_{,\Phi}^2}\right)\,, \label{Geff2}
	\end{equation}
	so that (\ref{per2}) takes the familiar form 
	\begin{eqnarray}
	\ddot{\delta}_{\rm m}+2H\dot{\delta}_{\rm m}-4\pi G_{\rm eff}\rho_{\rm m}\delta_{\rm m}=0. \;\label{per1}
	\end{eqnarray}
	Equation (\ref{per1}) is scale independent (as in the GR case) and the Newtonian gravitational constant  $G_N=8\pi G$ is replaced by the effective gravitational constant $G_{\rm eff}$ as given by equation (\ref{Geff2}): the scalar field perturbations $\delta\Phi$ showed themself in the definition of $G_{\rm eff}$. 
	In other words, beside causing the acceleration of the  expansion rate of the Universe, quintessence models can change the formation rate of structures.

	It is convenient to express equation \eqref{per1} in terms of the scale factor:
	\begin{equation}
	\delta_{m}^{''}+ \left (\frac{3}{a}+\frac{H'}{H} \right)\delta_{m}^{'}-\frac{3}{2a^2 F}\left(\frac{2F+4F_{,\Phi}^2}{2F+3F_{,\Phi}^2}\right)\Omega_{m}(a)\delta_{m}=0\label{ddm1},
	\end{equation}
	which we solve numerically setting an initial contrast of $\delta_{mi} = 5\times  10 ^{-5}$ and an initial scale factor of $a_i = 0.0001$, meaning that we are deep enough in the early matter dominated era. We verified that matter perturbations always stay in the linear regime. 
	
		\begin{table*}
		\centering
		\caption{Marginalized constraints (1$\sigma$ uncertainties) using the background observables of equation \eqref{eq:like-tot_chi} (top table) and the background and perturbation observables of equation \eqref{eq:like-tot_chi2} (bottom table). The last two columns show derived quantities relative to the central values.\label{tab:res}}
		\begin{tabular}{|c | c|  c|c| c|  c|c|}
			\hline \hline
			Model&$\Omega_{m0}$&$H_0$& $\bar{k}\times 10^4$&$\alpha$&${\xi}$\\
			\hline
			MSF&$ 0.303^{+0.004}_{-0.005}$&$ 69.12^{+0.47}_{-0.47}$&  $3.76^{+0.13}_{-0.15}$&$0.039^{+0.02}_{-0.03}$& -\\
			\hline
			NSF&$0.308^{+0.005}_{-0.005}$ & $68.44^{+0.53}_{-0.53}$& $3.5^{+0.24}_{-0.24}$&$0.07^{+0.05}_{-0.07}$& $-0.06^{+0.12}_{-0.08}$	\\
			\hline
			$\Lambda$CDM &$0.323^{+0.005}_{-0.005}$ & $67.15^{+0.38}_{-0.38}$&$-$&$-$&$-$	\\
			\hline \hline
		\end{tabular}
		\begin{tabular}{|c | c|  c|c| c|  c|c|}
			\hline \hline
			Model&$\Omega_{m0}$&$H_0$& $\bar{k}\times 10^4$&$\alpha$&${\xi}$&$\sigma_{80}$\\
			\hline
			MSF&$ 0.301^{+0.004}_{-0.004}$& $69.36^{+0.38}_{-0.43}$& $ 3.86^{+0.08}_{-0.08}$& $ 0.023^{+0.01}_{-0.02}$&$-$&$0.765^{+0.03}_{-0.03}$\\
			\hline	
			NSF&$0.300 ^{+0.002}_{-0.004}$& $69.74^{+0.32}_{-0.32}$&$ 3.87^{+0.05}_{-0.07}$&  $0.016^{+0.01}_{-0.01}$& $-0.06^{+0.18}_{-0.17}$&$0.763^{+0.03}_{-0.03}$\\
			\hline
			$\Lambda$CDM 
			&$0.323^{+0.005}_{-0.005}$& $67.17^{+0.4}_{-0.4}$&-&-&-&$0.745^{+0.03}_{-0.03}$\\
			\hline \hline
		\end{tabular}
	\end{table*}
		
	\begin{table}
		\centering
		\caption{Model selection results when using the background observables of equation \eqref{eq:like-tot_chi} (top table) and the background and perturbation observables of equation \eqref{eq:like-tot_chi2} (bottom table).\label{tabaic}}
		\begin{tabular}{|c | c| c|  c|c|c|}
			\hline \hline
			Model&$\chi^2_{\rm min}$&$\chi^2_{\rm red}$ &$\Delta \chi^2_{\rm min}$  &$\Delta$AIC&$\Delta$BIC\\
			\hline
			MSF&94.7&1.17&$-1.7$& $2.8$& $7.3$\\
			\hline
			NSF&95.6&$1.18$&-0.8&5.9 & $12.5$\\
			\hline
			$\Lambda$CDM&96.4&1.16&$-$&$-$&$-$\\
			\hline \hline
		\end{tabular}
		\begin{tabular}{|c | c| c|  c|c|c|}
			\hline \hline
			Model&$\chi^2_{\rm min}$&$\chi^2_{\rm red}$&$\Delta \chi^2_{\rm min}$&$\Delta$AIC&$\Delta$BIC\\
			\hline
			MSF&110.3&$1.13$&-0.8&3.7 & $8.5$\\
			\hline
			NSF&108.8&$1.12$&-2.3&4.46 & $11.6$\\
			\hline
			$\Lambda$CDM&111.1&1.11&$-$&$-$&$-$\\
			\hline \hline
		\end{tabular}
	\end{table}	
	
	\begin{figure}
		\begin{center}
			\includegraphics[width=\columnwidth]{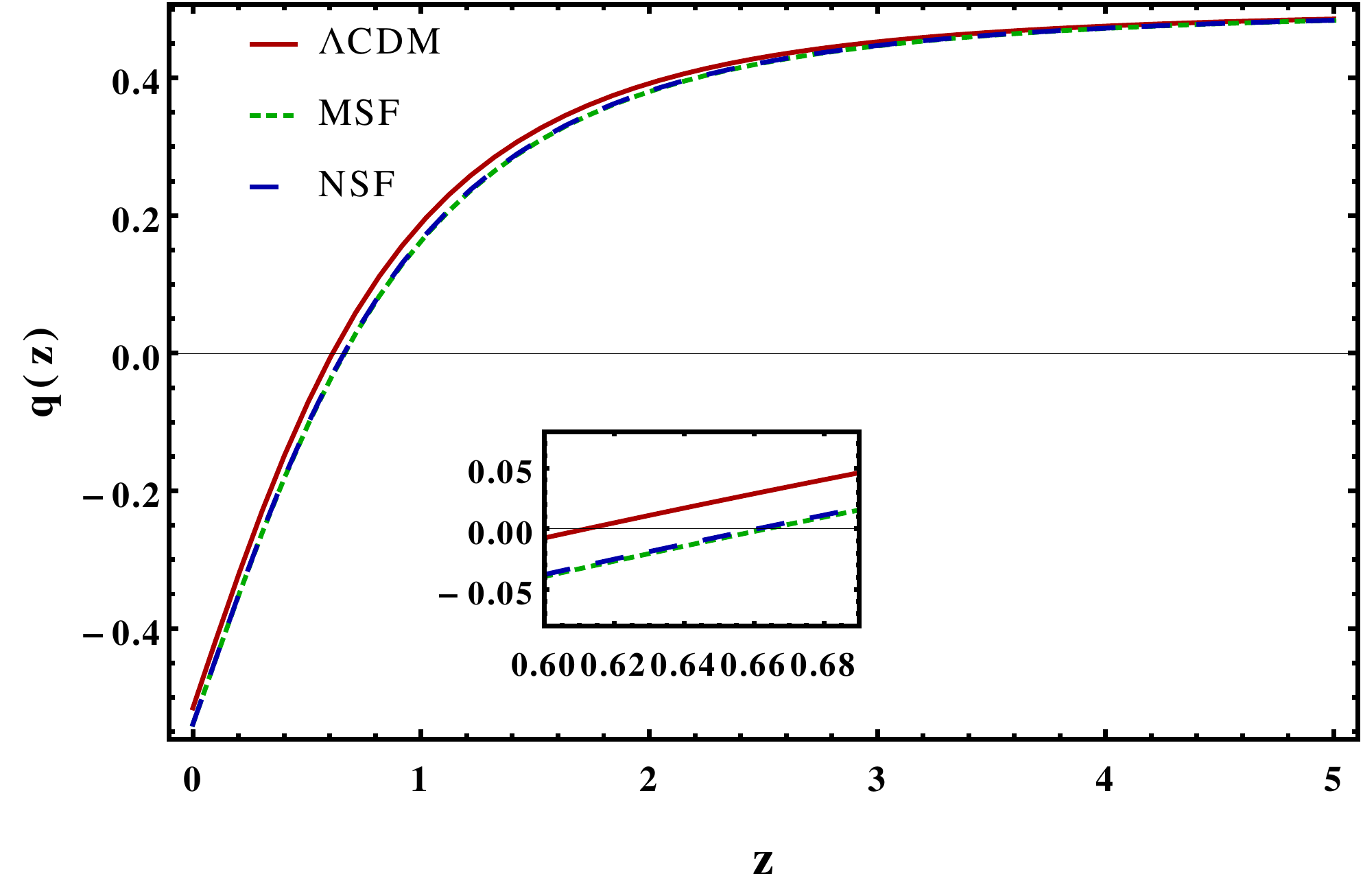}
			\caption{The evolution of the deceleration parameter $q$ for the best fit models given in the lower half of table~\ref{tab:res}.}\label{fig:q}
		\end{center} 
	\end{figure}
\begin{figure*}
	\begin{center}
		\includegraphics[width=14cm]{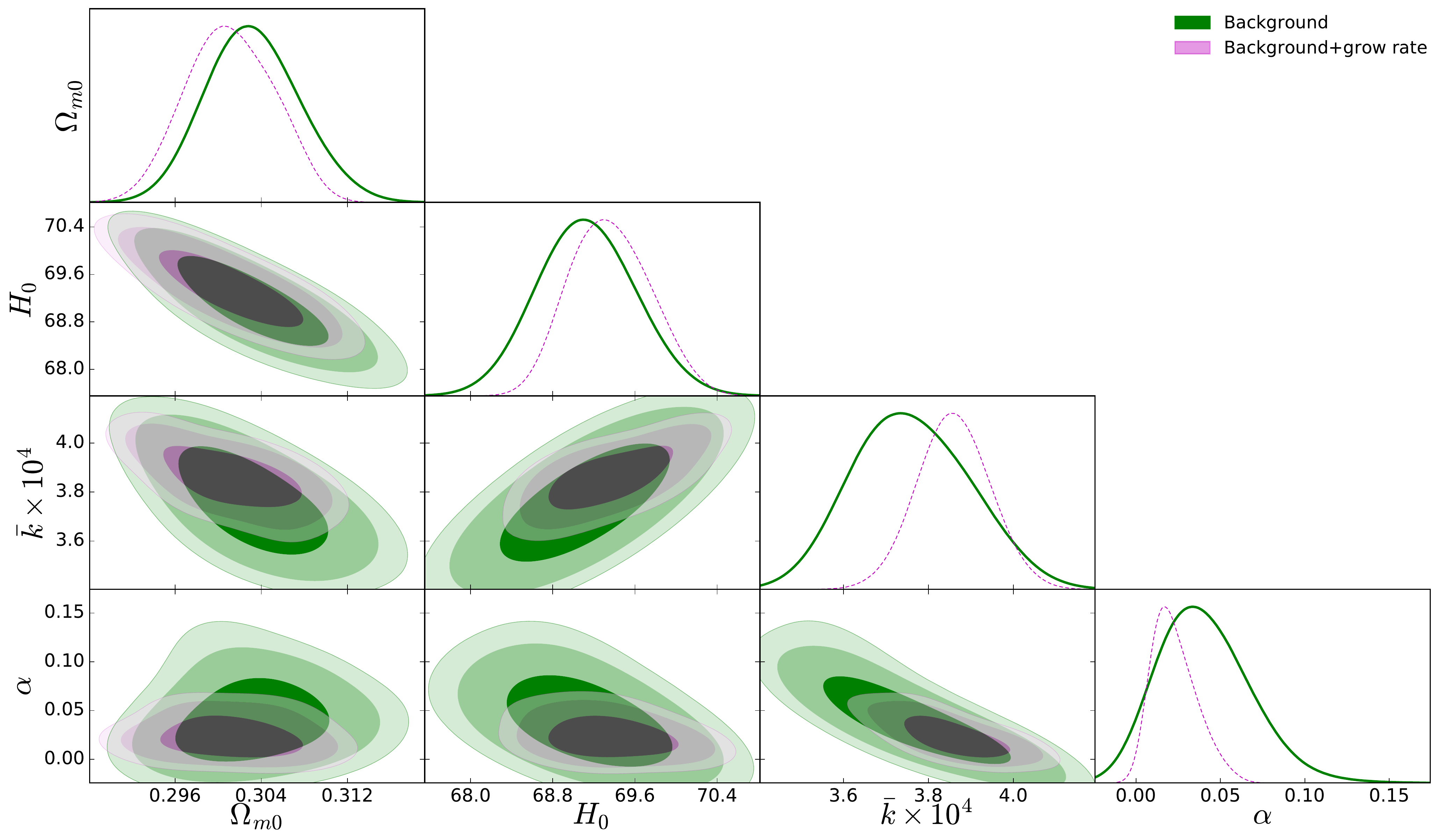}
		\includegraphics[width=14cm]{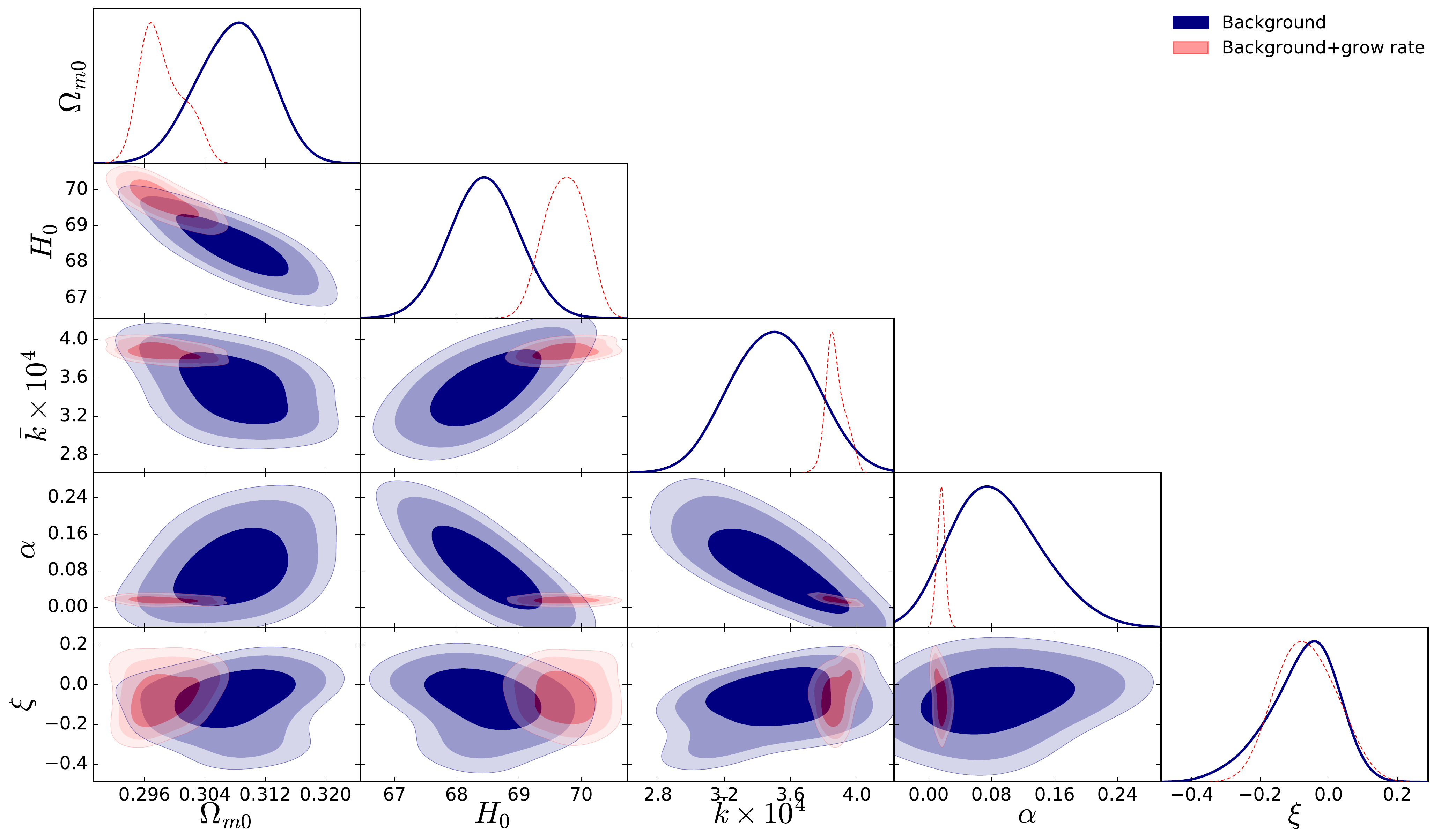}
		\caption{Marginalized 1-, 2- and 3$\sigma$ constraints on the MSF model (top panel) and NSF model (bottom panel) using both background (see equation \eqref{eq:like-tot_chi}) and background + perturbation data (see equation \eqref{eq:like-tot_chi2}). See table \ref{tab:res} for the numerical values.}
		\label{figc1}
	\end{center}
\end{figure*}

	In order to compare these models with observational data we calculate 
	the so-called linear growth rate, namely the logarithmic derivative of the linear density contrast with respect to the scale factor:
\begin{equation}
f(a)=\frac{d\ln \delta_m}{d\ln a} \,.
\end{equation}
The growth rate has been measured in several surveys at redshifts ranging from $z = 0.02$ up to $z = 1.4$.
	Another related quantity is the $\gamma$-index of matter perturbations, defined through $f(z)\simeq\Omega_m(z)^{\gamma(z)}$. The growth index can obviously be obtained according to \citep{Linder:2004ng, Linder:2007hg, Basilakos:2013ij}
	\begin{equation}
	\gamma(z)\cong \frac{\ln f(z)}{\ln \Omega_m(z)}.
	\end{equation}
\begin{table*}
	\centering
	\caption{As table \ref{tab:res} but also considering the local determination of $H_0$ by \citet{Riess:2019cxk}.\label{tab:res2}}
	\begin{tabular}{|c | c|  c|c| c|  c|c|}
		\hline \hline
		Model&$\Omega_{m0}$&$H_0$& $\bar{k}\times 10^4$&$\alpha$&${\xi}$\\
		\hline
		MSF&$ 0.302^{+0.005}_{-0.005}$&$ 69.23^{+0.49}_{-0.49}$&  $3.75^{+0.15}_{-0.15}$&$0.051^{+0.02}_{-0.04}$& -\\
		\hline
		NSF&$0.301^{+0.005}_{-0.005}$ & $69.37^{+0.45}_{-0.45}$& $3.89^{+0.01}_{-0.01}$&$0.03^{+0.01}_{-0.03}$& $0.03^{+0.12}_{-0.12}$	\\
		\hline
		$\Lambda$CDM &$0.318^{+0.005}_{-0.004}$ & $67.50^{+0.31}_{-0.37}$&$-$&$-$&$-$& 	\\
		\hline \hline
	\end{tabular}
	\begin{tabular}{|c | c|  c|c| c|  c|c| c|}
		\hline \hline
		Model&$\Omega_{m0}$&$H_0$& $\bar{k}\times 10^4$&$\alpha$&${\xi}$&$\sigma_{80}$ & $S_8$\\
		\hline	
		MSF&$ 0.299^{+0.003}_{-0.003}$& $69.63^{+0.3}_{-0.3}$& $ 3.90^{+0.05}_{-0.08}$& $ 0.02^{+0.01}_{-0.01}$&$-$&$0.769^{+0.04}_{-0.03}$ & $0.767^{+0.039}_{-0.030}$\\
		\hline
		NSF&$0.301 ^{+0.004}_{-0.004}$& $69.23^{+0.3}_{-0.3}$&$ 3.86^{+0.09}_{-0.13}$&  $0.021^{+0.011}_{-0.016}$& $-0.06^{+0.19}_{-0.19}$&$0.762^{+0.03}_{-0.03}$ & $0.764^{+0.033}_{-0.033}$\\
		\hline
		$\Lambda$CDM 
		&$0.318^{+0.005}_{-0.005}$& $68.09^{+0.40}_{-0.45}$&-&-&-&$0.748^{+0.03}_{-0.03}$ & $0.772^{+0.021}_{-0.021}$ \\
		\hline \hline
	\end{tabular}
\end{table*}

\begin{table}
	\centering
	\caption{As table \ref{tabaic} but also considering the local determination of $H_0$ by \citet{Riess:2019cxk}.\label{tabaic2}}
	\begin{tabular}{|c | c| c|  c|c|c|}
		\hline \hline
		Model&$\chi^2_{\rm min}$&$\chi^2_{\rm red}$ &$\Delta \chi^2_{\rm min}$  &$\Delta$AIC&$\Delta$BIC\\
		\hline
		MSF&104.7&1.3&-6.8& -2.3& $2.1$\\
		\hline
		NSF&107.5&$1.33$&-4.0&2.8 & $9.4$\\
		\hline
		$\Lambda$CDM&111.5&1.34&$-$&$-$&$-$\\
		\hline \hline
	\end{tabular}
	\begin{tabular}{|c | c| c|  c|c|c|}
		\hline \hline
		Model&$\chi^2_{\rm min}$&$\chi^2_{\rm red}$&$\Delta \chi^2_{\rm min}$&$\Delta$AIC&$\Delta$BIC\\
		\hline
		MSF&119.8&1.22& -12.8&-8.3& $-3.5$\\
		\hline
		NSF&121.0&$1.23$&-11.6&-4.8 & $2.4$\\
		\hline
		$\Lambda$CDM&132.6&1.33&$-$&$-$&$-$\\
		\hline \hline
	\end{tabular}
\end{table}

	In the left top panel of  figure~\ref{fig:grow} we show the linear growth factor $f(z)$ for the different cosmological models studied in this work. In this figure the best-fit values from the lower half of table~\ref{tab:res} are adopted.
	This panel shows that DE decreases the amplitude of matter perturbations at low redshifts. Notice that at high redshifts the influence of DE on the growth of perturbations is negligible and consequently the growth function goes to unity, which corresponds to the matter dominated Universe. We observe that the suppression of the amplitude of matter fluctuations in scalar tensor cosmologies starts sooner as compared with the concordance $\Lambda$CDM model. In the right top panel of  figure~\ref{fig:grow} we see the evolution of $\sigma_8(z)$ as a function of redshift $z$.
	$\sigma_8(z) $ is the redshift-dependent rms fluctuations of the linear density field within spheres of radius $R = 8h^{-1}$Mpc.
	It shows that $\sigma_8(z)$ in ST models  is
	larger than the one in the case of the $\Lambda$CDM universe.
The right bottom panel of the figure~\ref{fig:grow} shows the corresponding growth indexes which are larger for the $\Lambda$ cosmology with respect to the minimally and non-minimally coupled models at high redshifts and they close to each other at small redshifts ($z<1$).
	A robust measurable quantity in redshift surveys is $f(z)\sigma_8(z) $.
	This quantity is shown in the bottom panel of figure~\ref{fig:grow}, together with the observational data to be discussed in the next Section.

	Another quantity that is interesting in the study of DE models is the cosmological deceleration-acceleration transition redshift, $z_t$, which is defined as
	the redshift at which $\ddot{a} = 0$. In figure~\ref{fig:q}, we show the redshift evolution of the deceleration parameter $q=-\ddot{a}/aH^2$ for the best fit models given in the lower half of table~\ref{tab:res}. The transition point for the NSF model is $z_t = 0.661$, which is near the  MSF one ($z_{t} = 0.664$) and it is $z_t=0.612$ for the $\Lambda$CDM model. As expected, at early enough times, $q$ tends to $1/2$ since the universe is matter dominated. The latter results are  in good agreement with the values obtained in \citet{Farooq:2016zwm}.

\begin{figure*}
	\begin{center}
		\includegraphics[width=8.7cm]{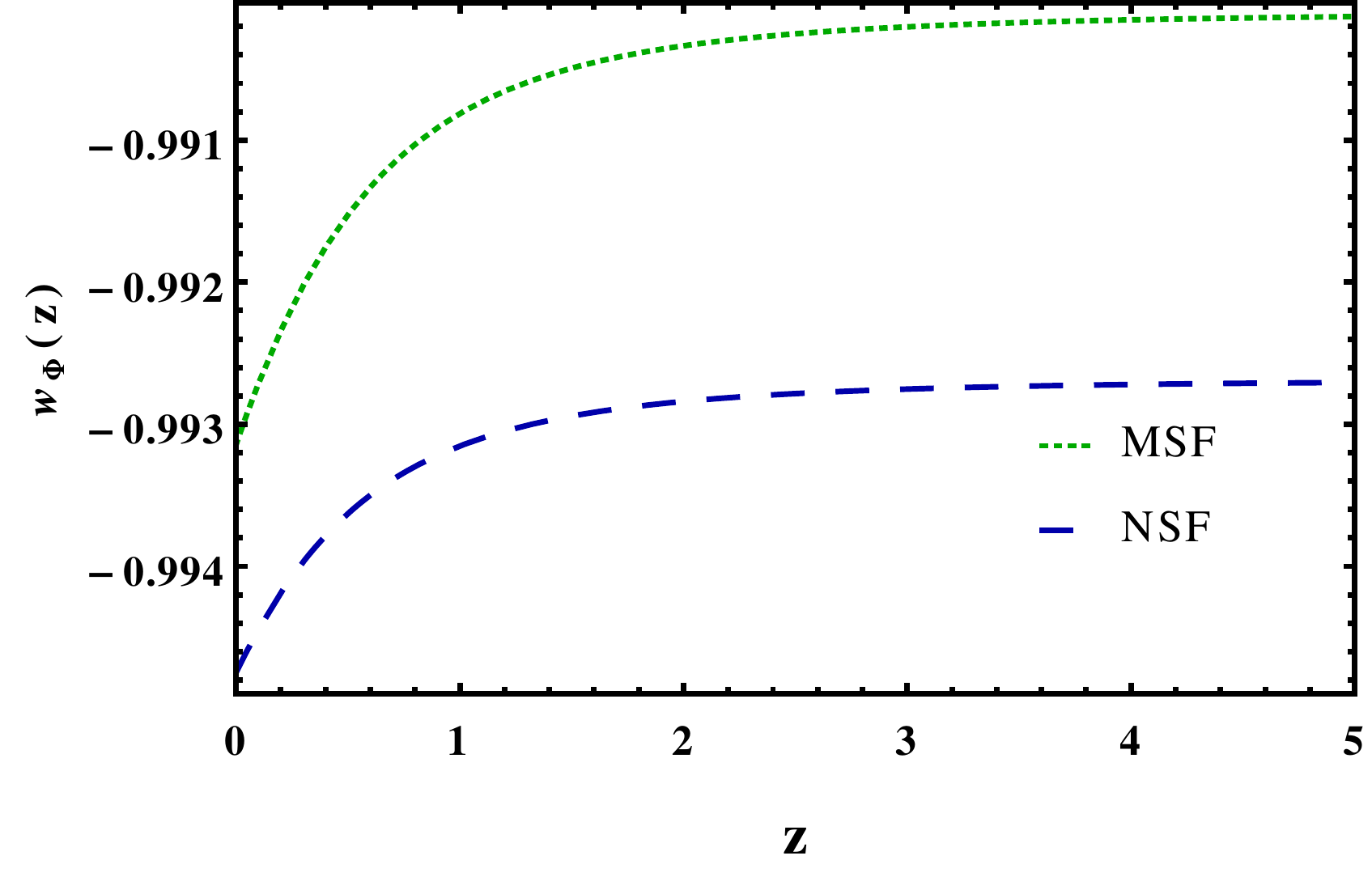}
		\includegraphics[width=8.7cm]{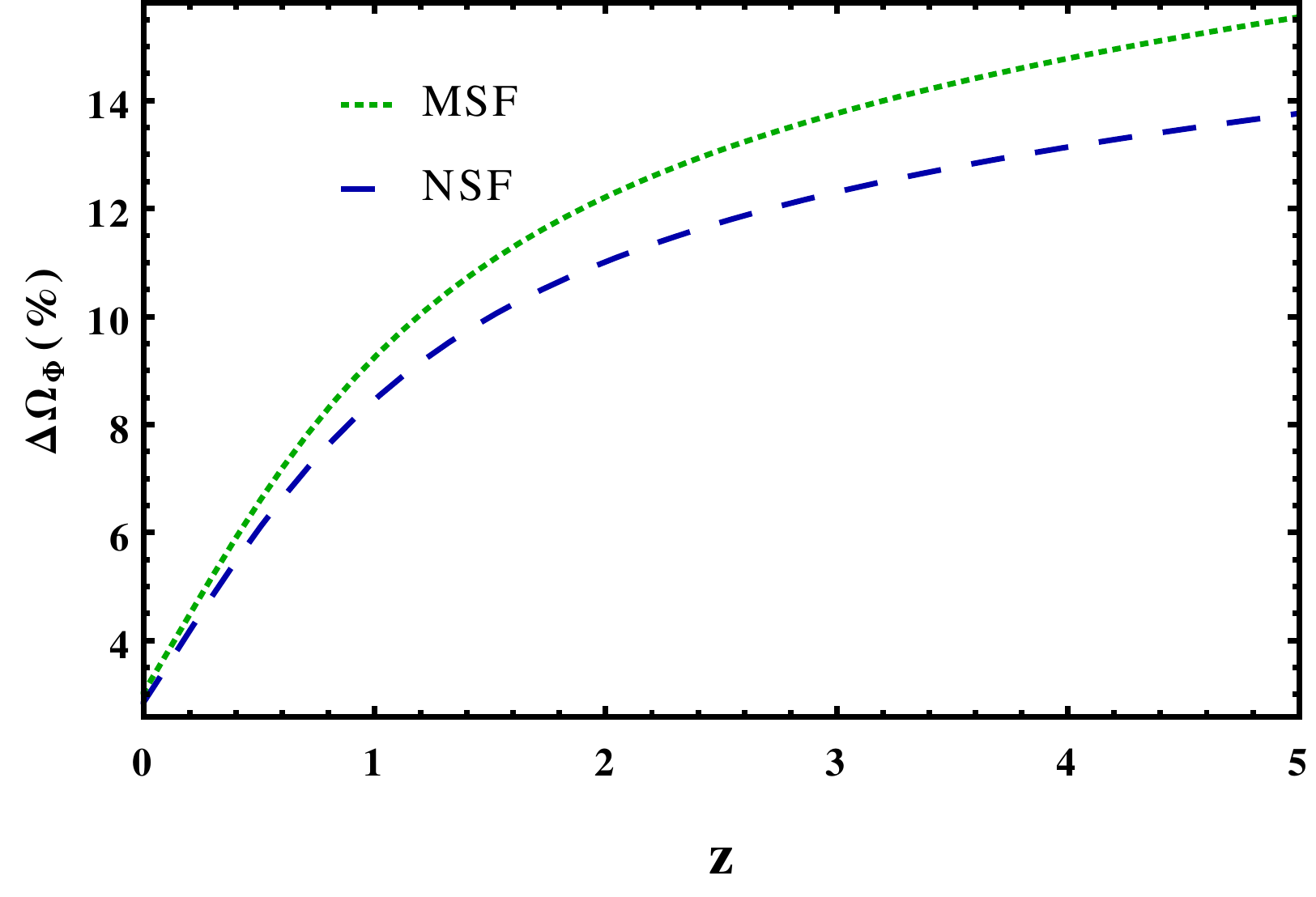}
		\includegraphics[width=8.7cm]{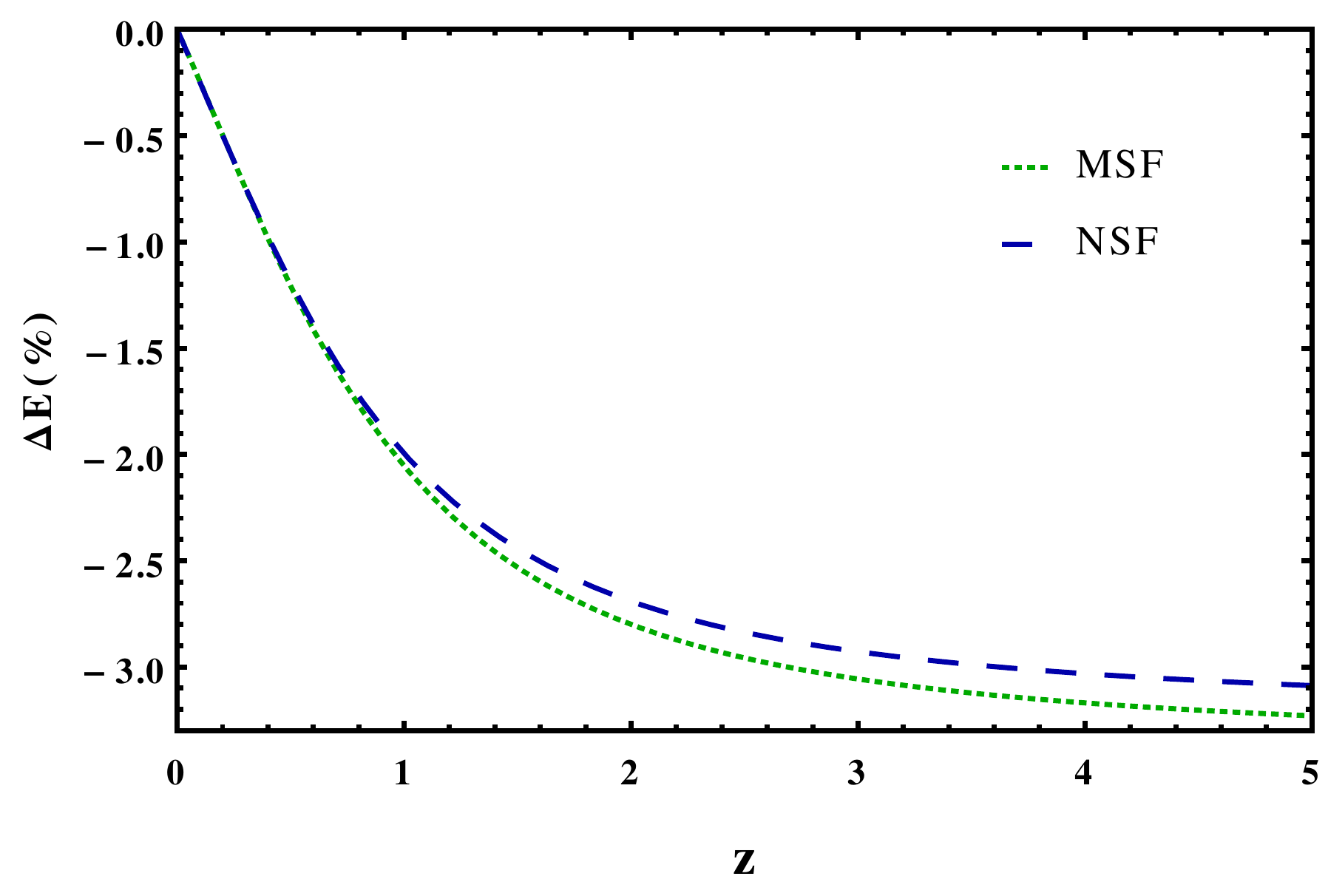}
		\includegraphics[width=8.7cm]{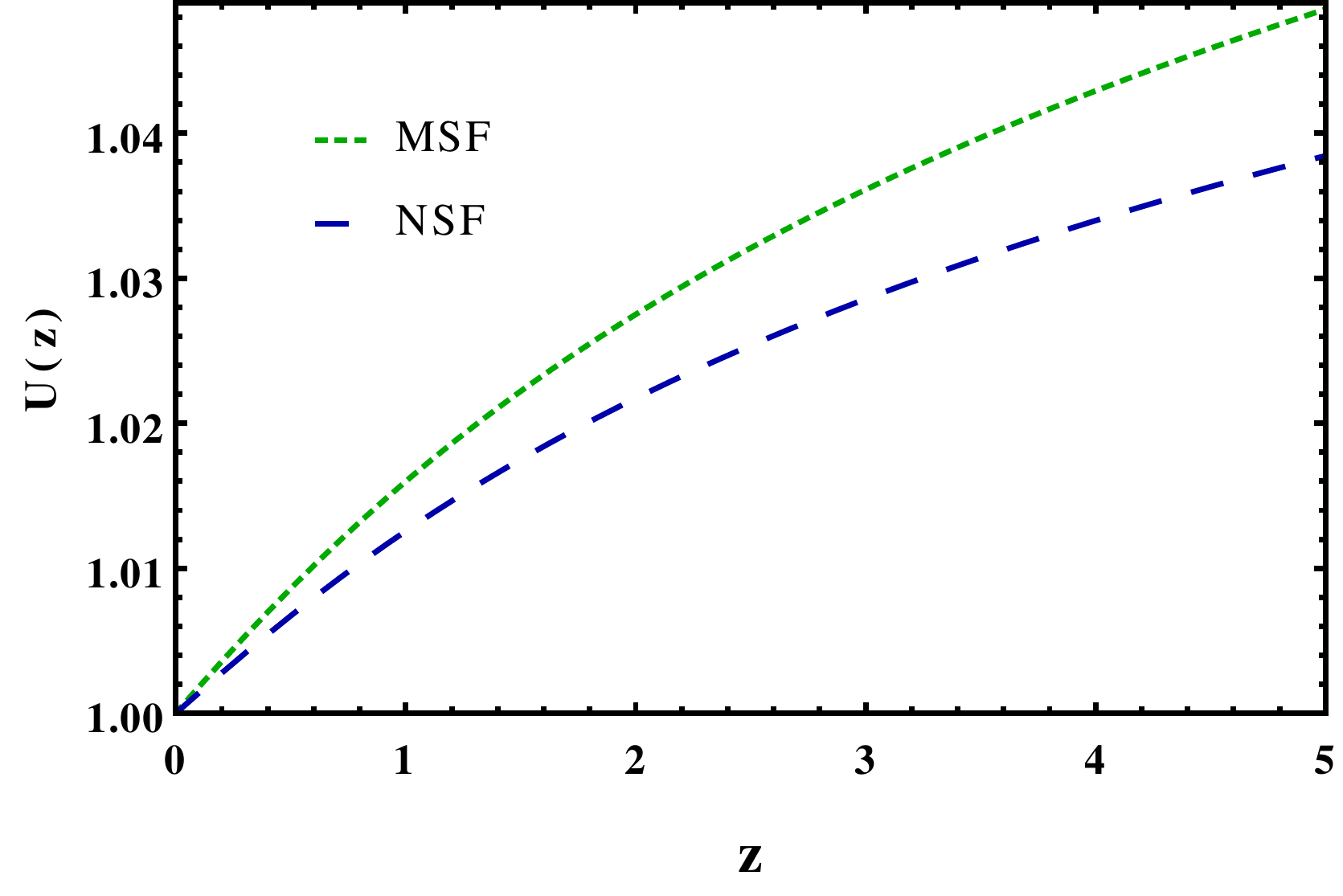}
		\caption{The redshift evolution of $w_{\Phi}(z)$ (top left panel), the relative deviations  $\Delta \Omega_{\Phi}(z)$ (top right panel) and $\Delta E(z)$ (bottom left panel) with respect to $\Lambda$CDM, and the Ratra-Peebles potential $U(\Phi)$ normalized to $U(a=1)$ (bottom right panel) for the MSF and NSF models. The green dotted curve indicates the minimal scalar field model (MSF), the  blue dashed curve corresponds to the non-minimal scalar field model (NSF).}\label{fig:twomodel}
	\end{center}
\end{figure*}

\begin{figure}
	\begin{center}
		\includegraphics[width=\columnwidth]{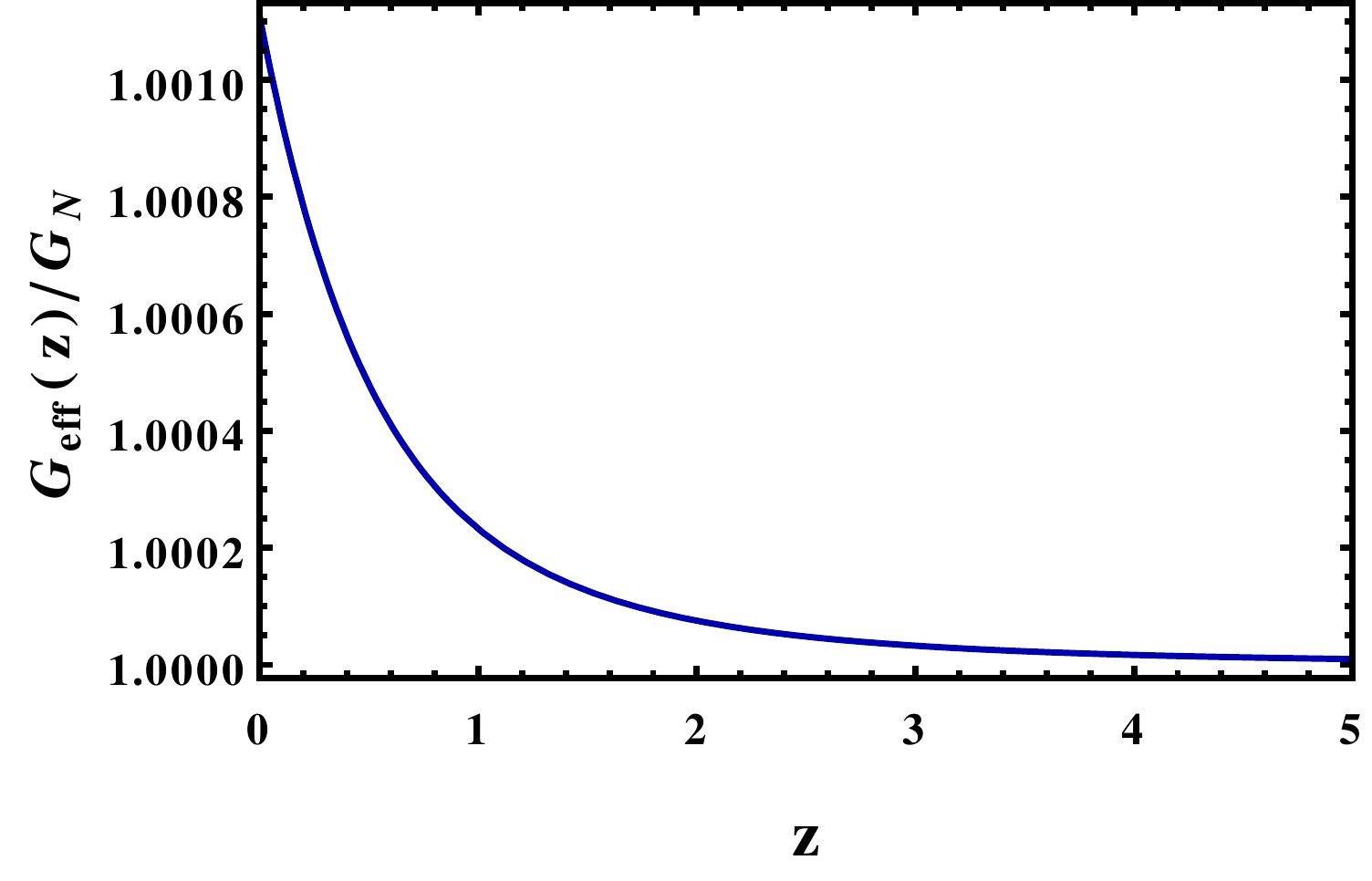}
		\caption{The redshift evolution of  the effective gravitational constant for the NSF best-fit model of the lower half of table \ref{tab:res}.}\label{figG}
	\end{center}
\end{figure}

\begin{figure*}
	\begin{center}
		\includegraphics[width=8cm]{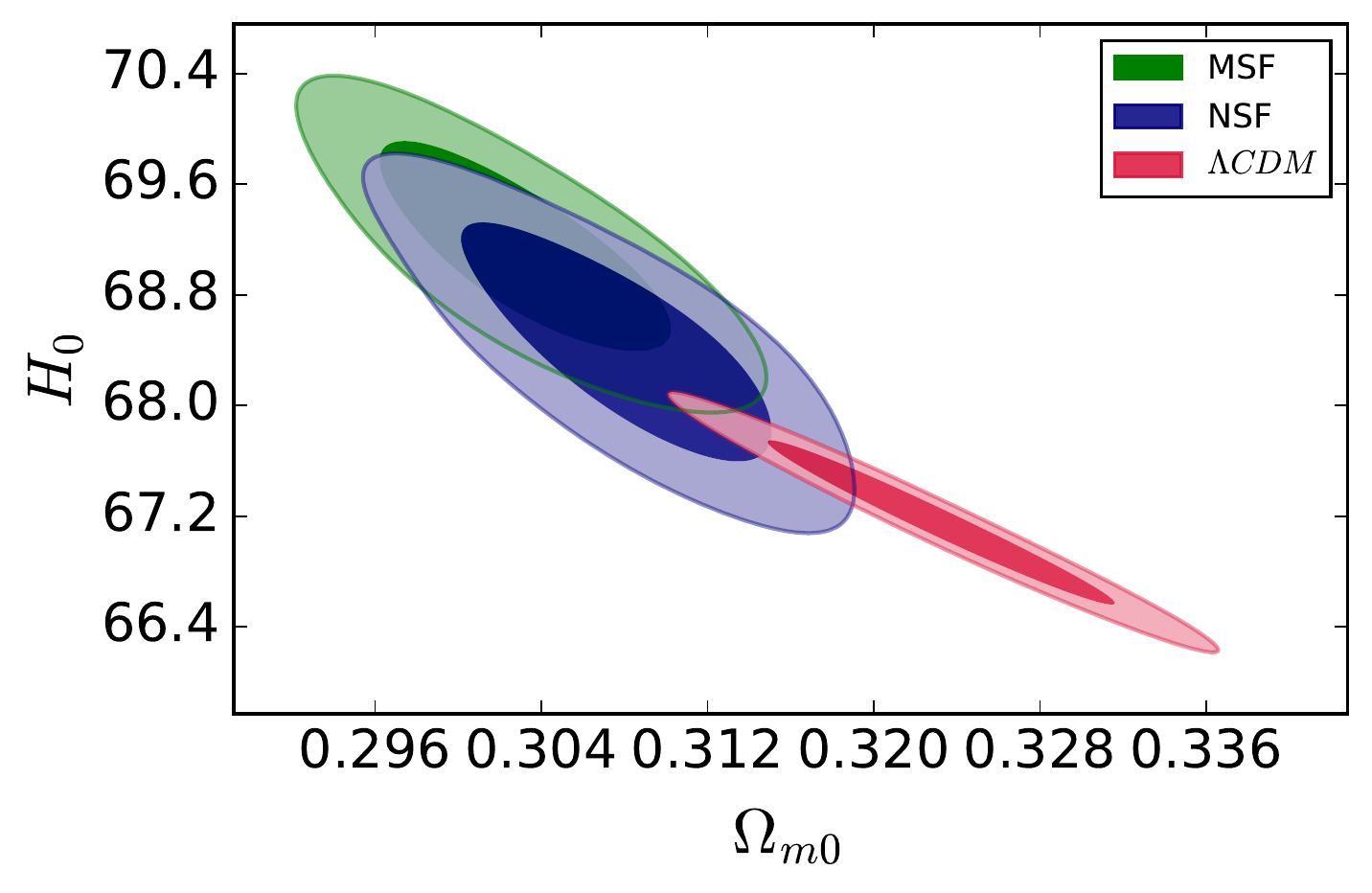}
		\includegraphics[width=8cm]{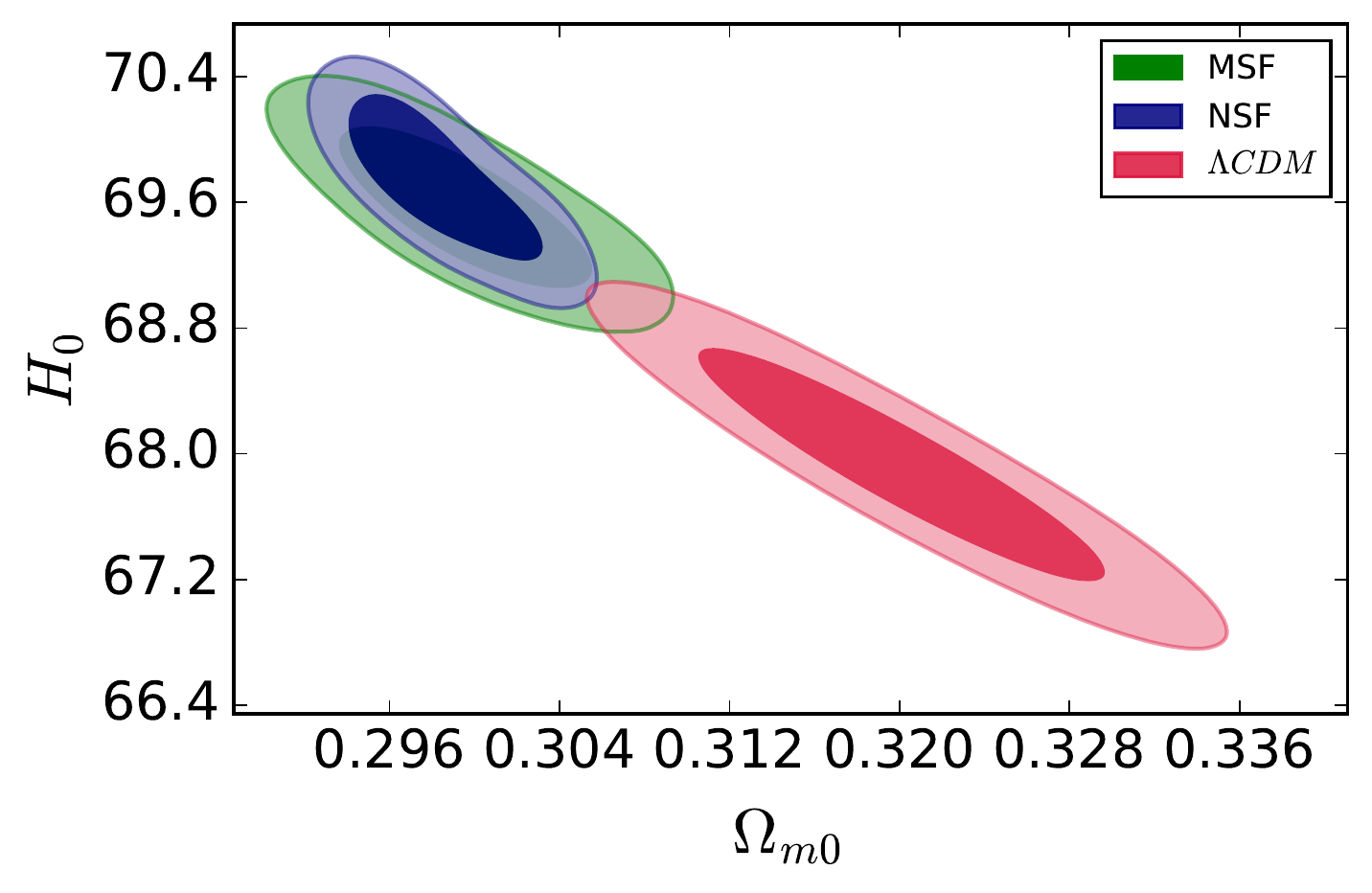}
		\includegraphics[width=8cm]{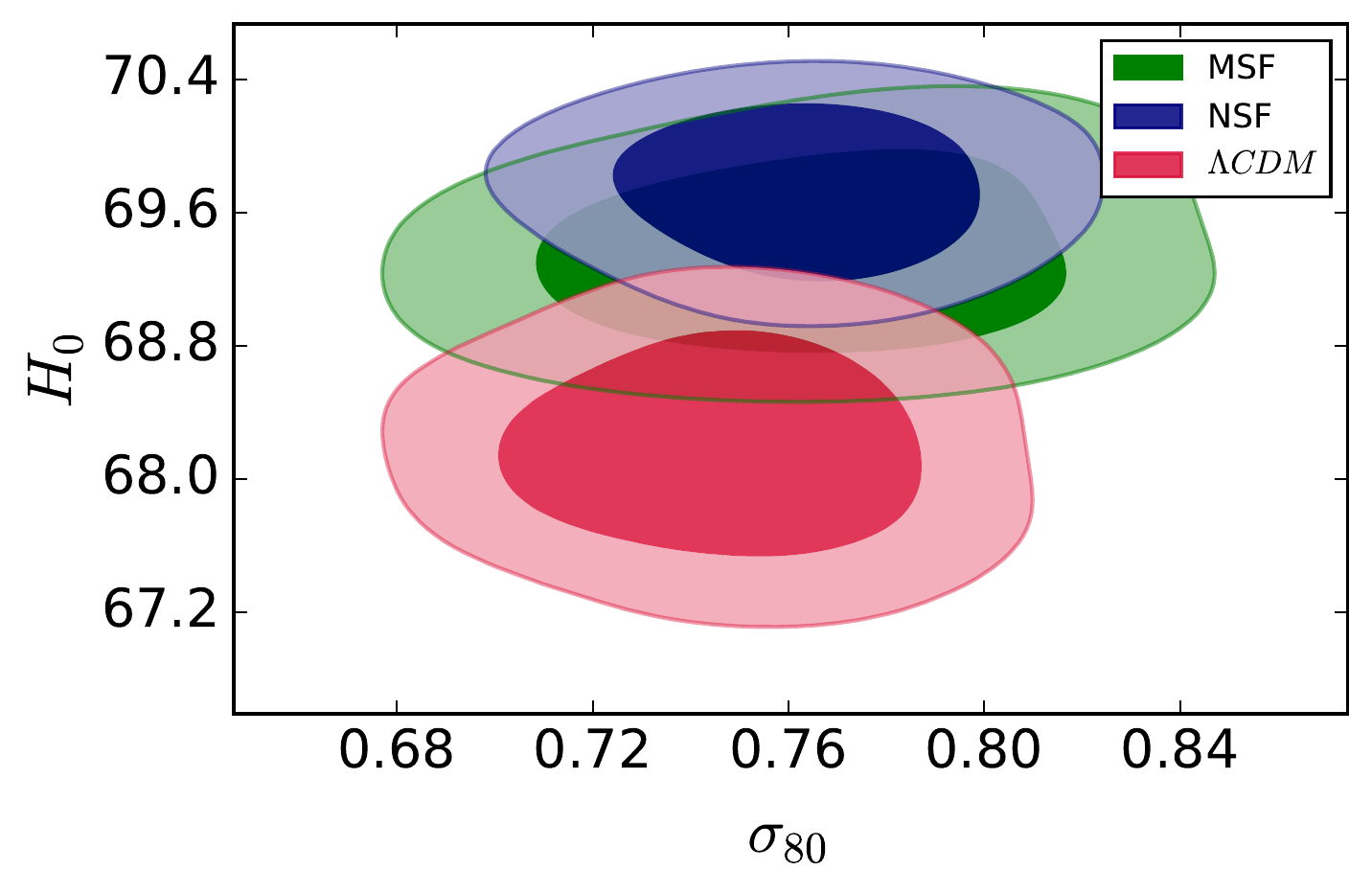}
		\includegraphics[width=8cm]{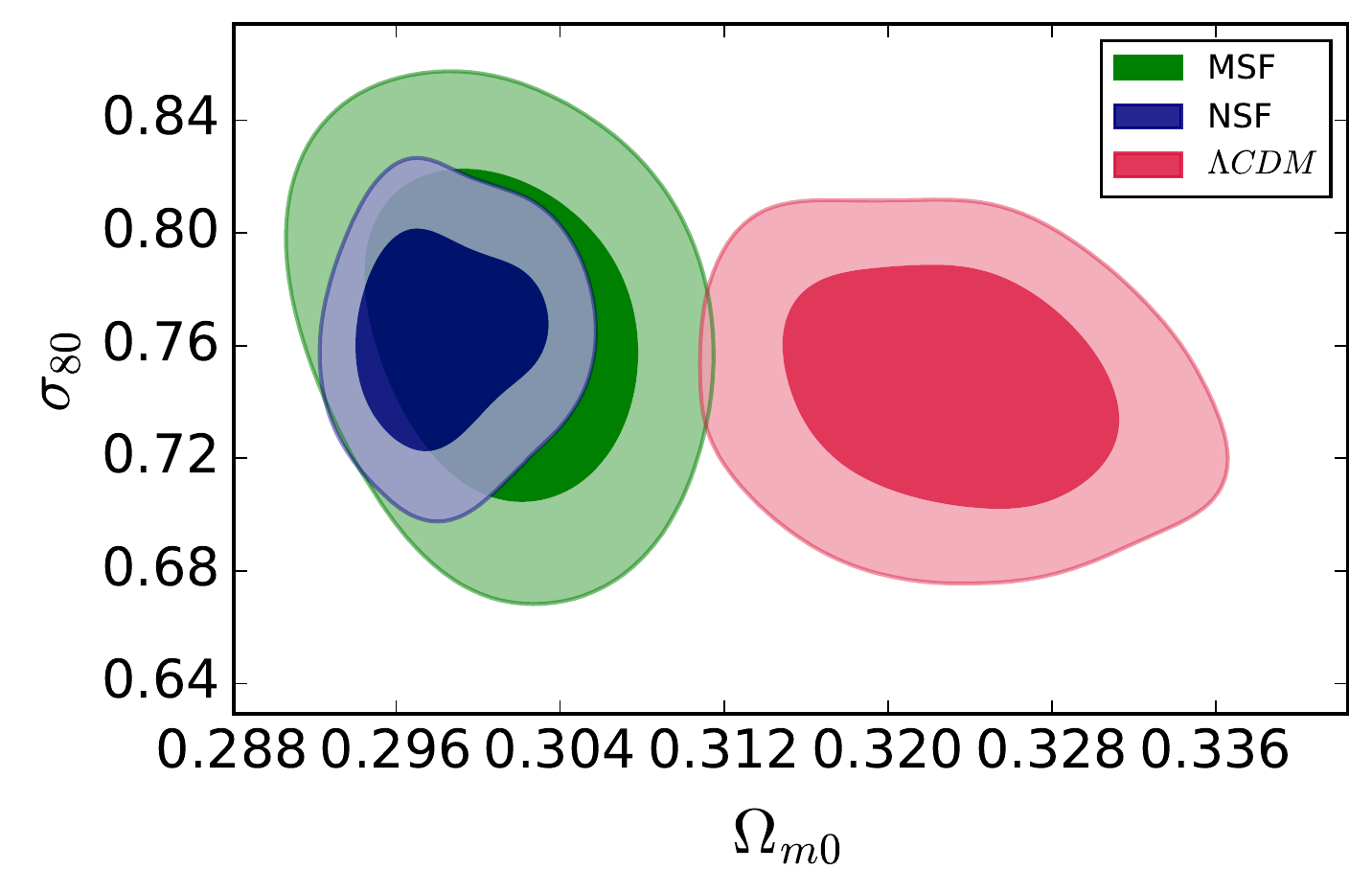}
		\caption{
			$1\sigma$ and $2\sigma$
			confidence levels for the parameters that are in common with the $\Lambda$CDM model.
			The top left panel is relative to background data only (see \eqref{eq:like-tot_chi}), while the other panels are relative to background and perturbation data (see \eqref{eq:like-tot_chi2}).}
		\label{figc2}
	\end{center}
\end{figure*} 
	\section{Scalar Field Models Versus Data}\label{sec3}
	In this section we wish to constrain the parameters of the MSF  and NSF models using the latest cosmological data at the background and perturbation level via the Markov Chain Monte Carlo (MCMC) method. The goal is to compare the MSF, NSF and $\Lambda$CDM models.

	We will perform two analyses. The first one will use a total of 86 data points from background observables, specifically  
	from type Ia supernovas (SNIa, 40 points from the Pantheon binned sample), baryon acoustic oscillations (BAO, 11 points),  cosmic chronometers ($H(z)$, 31 points), cosmic microwave background (CMB, 3 points) and big bang nucleosynthesis (BBN, 1 point). We briefly describe each dataset in Appendix~\ref{datasets}.
	The total likelihood is given by:
	\begin{equation}\label{eq:like-tot}
	{\cal L}_{\rm tot, 1}({\bf p})={\cal L}_{\rm sn} \times {\cal L}_{\rm bao} \times {\cal L}_{\rm cmb} \times {\cal L}_{h} \times
	{\cal L}_{\rm  bbn}  ,\;
	\end{equation}
	or (${\cal L} \simeq e^{-\frac{\chi^2}{2}}$):
	\begin{equation}\label{eq:like-tot_chi}
	\chi^2_{\rm tot,1}({\bf p})=\chi^2_{\rm sn}+\chi^2_{\rm bao}+\chi^2_{\rm cmb}+\chi^2_{h}+\chi^2_{\rm bbn}\;,
	\end{equation}
	where the statistical vector ${\bf p}$ is $\{\Omega_{b0}, \Omega_{dm0},H_0\}$, $\{\Omega_{b0}, \Omega_{dm0},H_0,\alpha,\bar{k}\}$ and $\{\Omega_{b0}, \Omega_{dm0},H_0,\alpha,\bar{k},\xi\}$ for $\Lambda$CDM, MSF and NSF models, respectively. Notice that $\Omega_{b0}$ and $\Omega_{dm0}$ are the current values of the energy densities of baryonic and dark matter, respectively. 
	 The value that minimizes $\chi^2_{\rm tot}$ (maximizes ${\cal L}_{\rm tot}$) defines the best-fit model and the corresponding $\chi^2_{\rm min}$.
	In the second analysis we also include perturbation observables, specifically 18 redshift space distortion (RSD) independent data points:
	\begin{equation}\label{eq:like-tot2}
	{\cal L}_{\rm tot, 2}({\bf p})={\cal L}_{\rm tot, 1}({\bf p}) \times {\cal L}_{\rm  fs}({\bf p}) ,\;
	\end{equation}
	or:
	\begin{equation}\label{eq:like-tot_chi2}
	\chi^2_{\rm tot,2}({\bf p})=\chi^2_{\rm tot,1}({\bf p}) +\chi^2_{\rm fs}({\bf p}) \;.
	\end{equation}
	In this analysis, ${\bf p}$ also includes $\sigma_{80}=\sigma_8(z=0)$ in addition to the parameters mentioned above.
	Comparing the two analyses will allow us to gauge the statistical strength of RSD data.

\subsection{Parameter estimation}

	Figures~\ref{figc1} summarizes the statistical analyses relative to equations \eqref{eq:like-tot_chi} and \eqref{eq:like-tot_chi2}, whose numerical results are listed in table~\ref{tab:res}.
	Note that we are showing the total matter density parameter $\Omega_{m0}=\Omega_{b0}+ \Omega_{dm0}$.
	The MSF, NSF and $\Lambda$CDM were analyzed.
It is clear that growth rate data significantly improve the constraints by breaking many degeneracies.
In particular, it is interesting to note the value of $H_0$ is larger as compared with $\Lambda$CDM, thus decreasing the tension with local determinations of $H_0$ (see \citet{Camarena:2018nbr} and references therein).
Also interesting is the fact that, for the NSF, the scalar potential steepness parameter $\alpha$ is constrained to be different from zero, signaling an effective equation of state different from $-1$.
The scalar potential scale $\bar k$ is different from zero because a nonzero DE is required in order to fit the data.

\begin{figure*}
	\begin{center}
		\includegraphics[width=14cm]{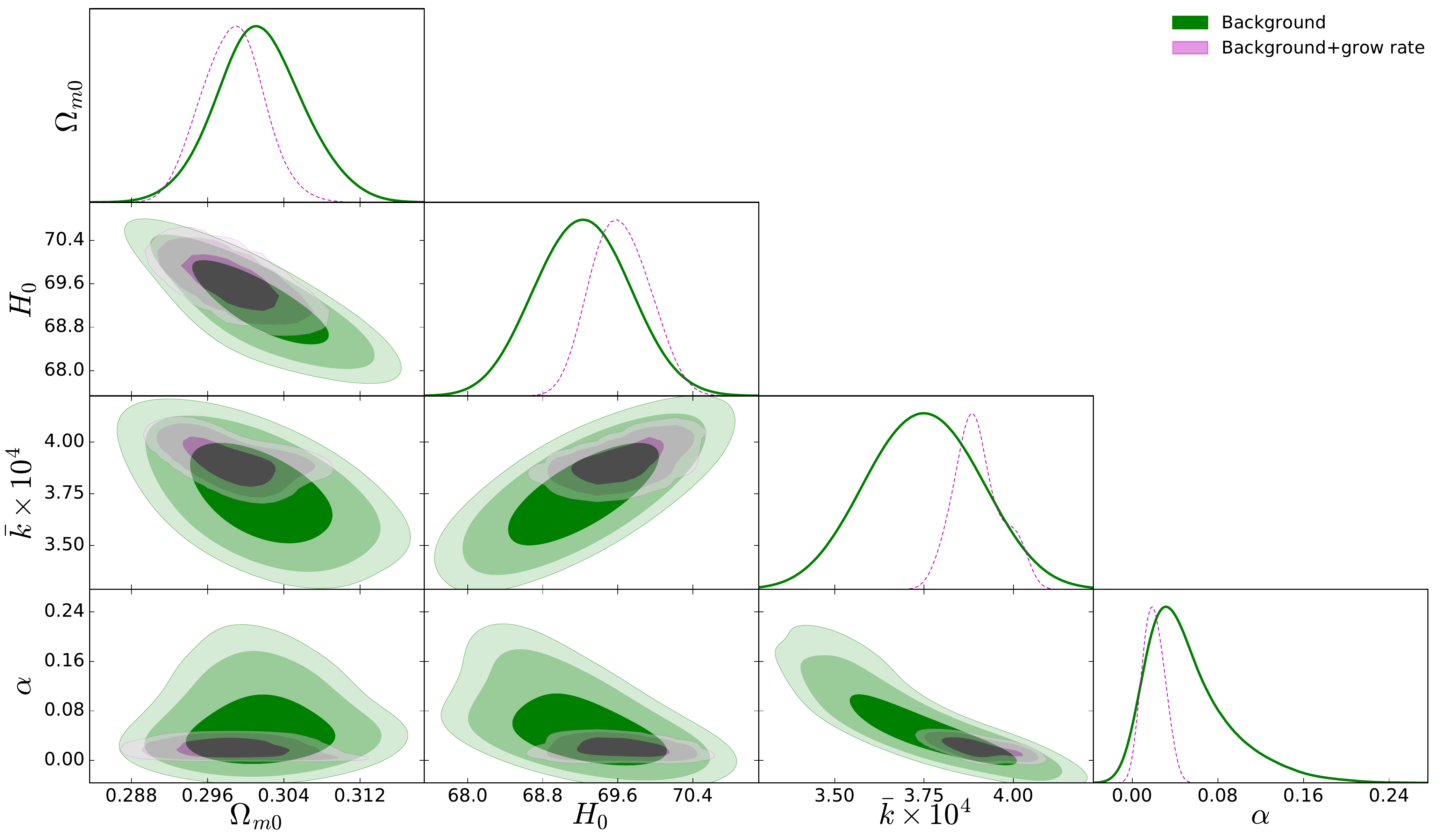}
		\includegraphics[width=14cm]{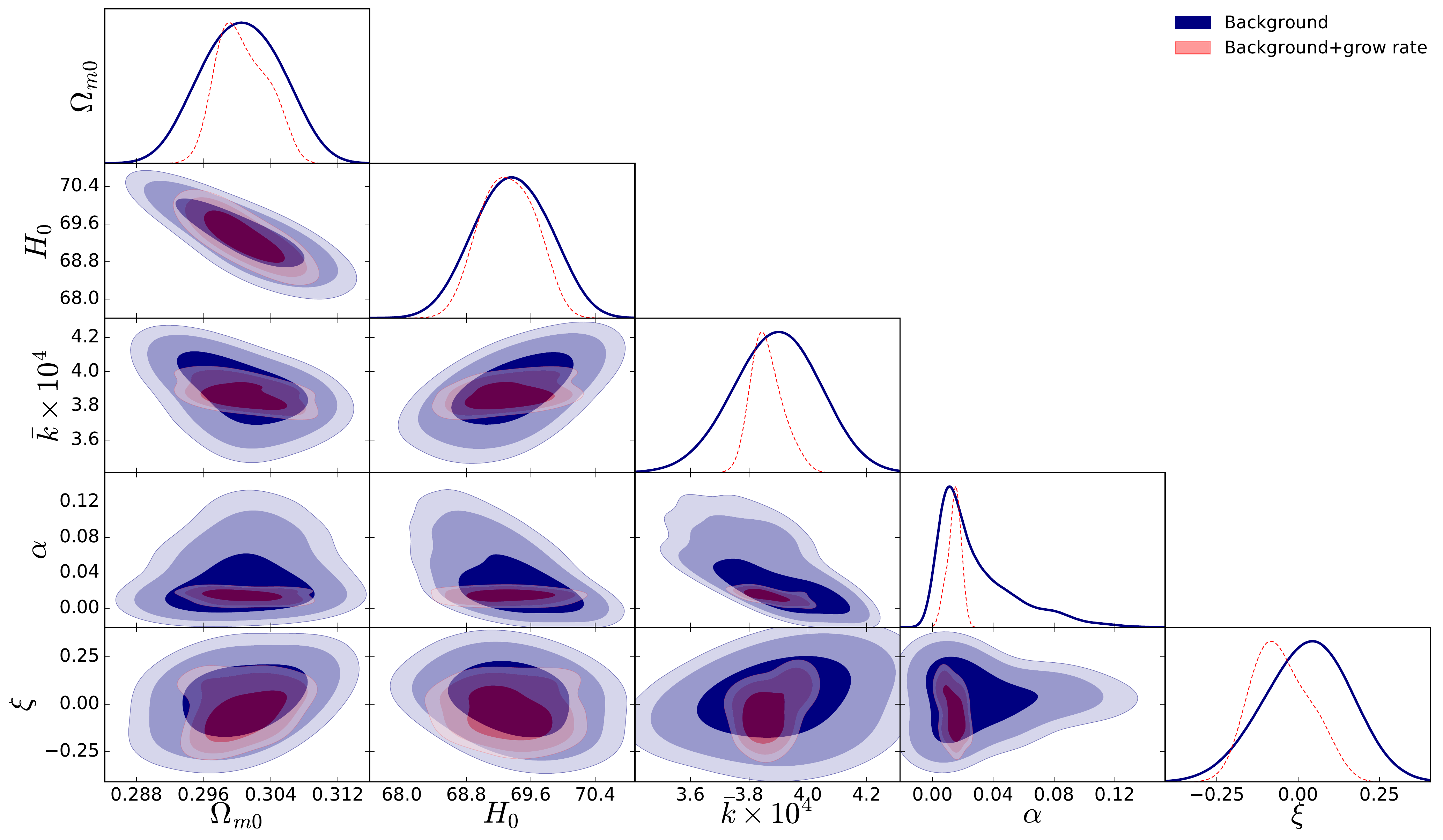}
		\caption{As figure \ref{figc1} but also considering the local determination of $H_0$ by \citet{Riess:2019cxk}.
			See table \ref{tab:res2} for the numerical values.}
		\label{figc22}
	\end{center}
\end{figure*}

\begin{figure}
	\begin{center}
		\includegraphics[width=\columnwidth]{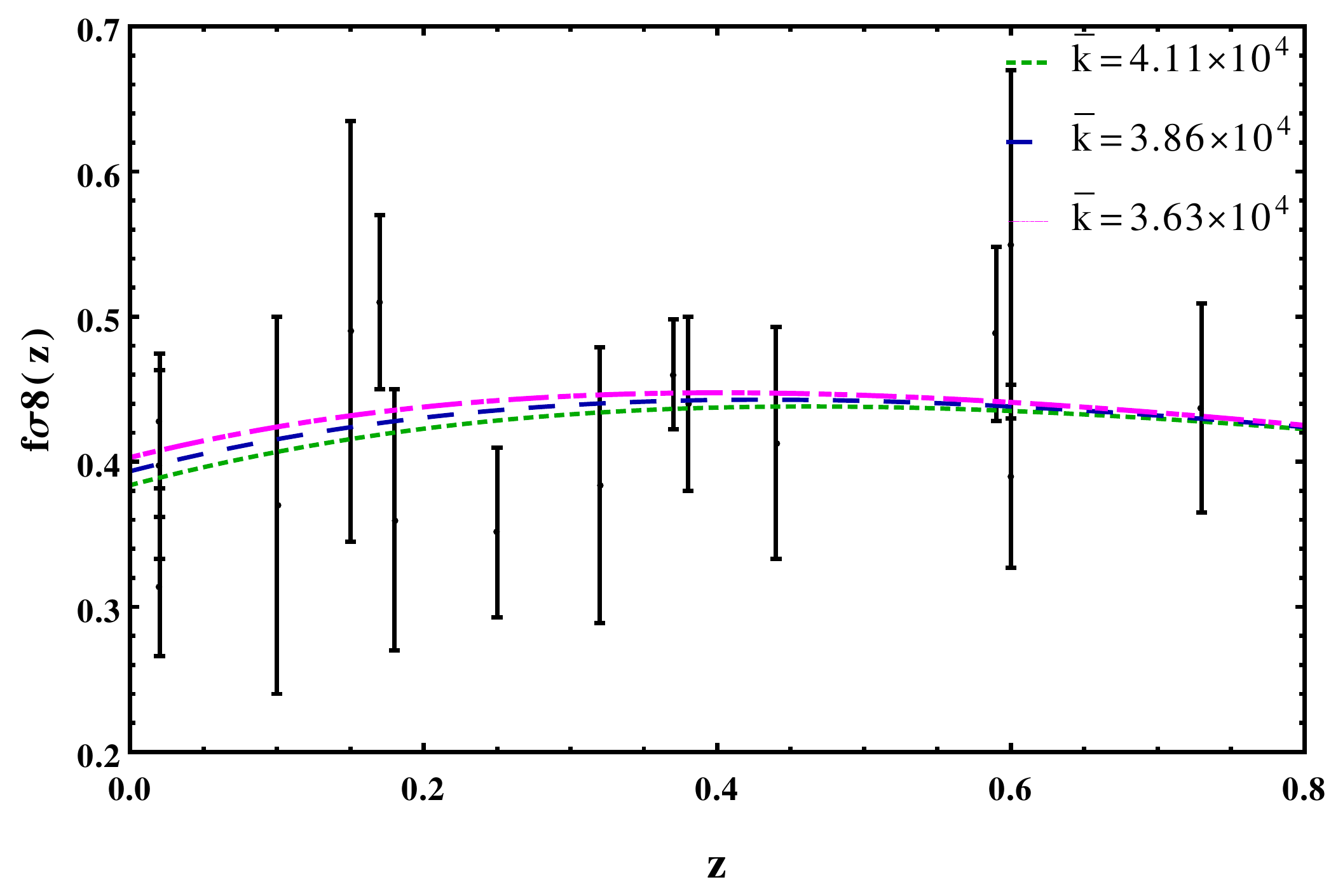}
		\caption{The redshift evolution of  the combination $f(z)\sigma_8(z)$ for the NSF model. The  dot-dashed, dashed and dotted curves correspond to ${\bar{k}}=4.11\times 10^4 , \alpha = 0.035 (+3\sigma)$, $\bar{k}_{\rm best}=3.86\times 10^4 , \alpha_{best} = 0.021$  and ${\bar{k}}=3.63\times 10^4 , \alpha = 0.001 (-3\sigma)$, respectively (for the other parameters we adopted the values given in table \ref{tab:res2}).}\label{kbar}
	\end{center}
\end{figure}

\begin{figure}
	\begin{center}
		\includegraphics[width=\columnwidth]{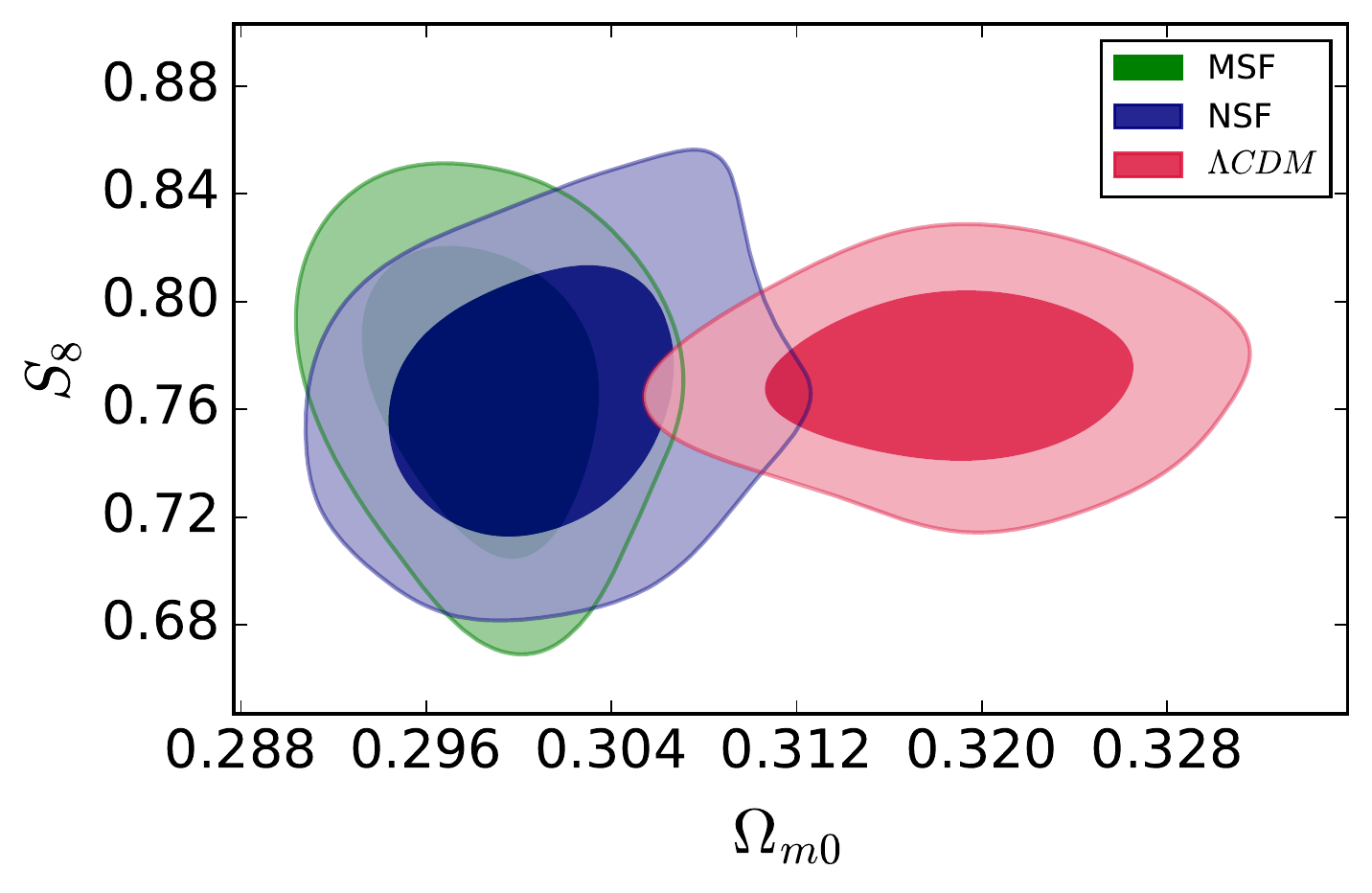}
		\caption{Cosmological constraints in the $S_8$ versus $\Omega_{m0}$ plane for MSF, NSF and $\Lambda$CDM cosmologies using background (including local $H_0$) and perturbations observables.  }\label{fig:S8}
	\end{center}
\end{figure}

\subsection{Best-fit models}

It is interesting to plot the evolution of the main cosmological quantities for the parameter values of table \ref{tab:res}. Although these values refer to the modes of the marginalized posteriors, their numerical value is very close to the actual best fit. In figure~\ref{fig:twomodel} we plotted $w_{\Phi}$ (top left panel),
the relative difference $\Delta \Omega_{\Phi}$ of  the energy density parameter of the scalar field  with respect to $\Omega_{\Lambda}$ (top right panel), $\Delta E$ (bottom left panel) and $U(\Phi)$ (bottom right panel), as a function of redshift $z$ for the different cosmological models presented in the lower half of table \ref{tab:res}. One can see that the scalar field is in the quintessence regime ($w_{\Phi}>-1$) and has a larger energy density as compared to $\Lambda$. We obtain the present values
	of the equation of state $w_{\Phi}= -0.993^{+0.005} _{-0.008}$ and  $w_{\Phi}=-0.995^{+0.0045}_{-0.005}$
	for the MSF and NSF models, respectively. Also, the
	current value of the function $F$  for the NSF model
	is $0.998^{+0.00004}_{-0.00004}$. Note that for MSF, we have $F=1$ at any redshift.
In figure~\ref{figG}, for the NSF model, we show  the redshift evaluation of the effective gravitational constant with respect to the Newtonian gravitational constant, $G_{\rm eff}(z)/G_N$. We obtain the value   
$G_{\rm eff} /G_N = 1.0011$
 at present time while at high redshifts it is $G_{\rm eff} /G_N = 1$  as in the case of minimally coupled quintessence models.

\subsection{Model selection}

The model with the lowest $\chi^2_{\rm min}$ is the one that best fits the data. However, it is not necessarily the model that should be selected as the best one as it could feature parameters that are poorly constrained by the data. A model with a poorly constrained (or ``useless'') parameter should be penalized.
The MSF model has 2 more parameters ($\alpha,\bar{k}$), while the NSF has three more parameters ($\alpha,\bar{k},\xi$).

In order to deal with model selection a standard approach is to compute the reduced $\chi^2$, the Akaike Information Criterion (AIC) and Bayesian Information Criterion (BIC):
\begin{align}
\chi^2_{\rm red} &= \frac{\chi^2_{\rm min}}{N-M} \,, \\
{\rm AIC}&=\chi^2_{\rm min}+2M+\frac{2M(M+1)}{N-M-1} \,, \label{aic}\\
{\rm BIC}&=\chi^2_{\rm min}+M\ln N \,, \label{bic}
\end{align}
where $N$  is the number of data points and $M$ the number of free parameters in the model.
A model with more parameters will have a negative $\Delta \chi^2_{\rm min}$ with respected to the (nested) model with less parameters. Roughly speaking, the practical effect of the terms depending on $N$ and $M$ in equations (\ref{aic}-\ref{bic}) is to make $\Delta \chi^2_{\rm min}$ less negative (favoring less the model with more parameters) and, depending on the case, making it even positive (favoring the model with less parameters).

In table~\ref{tabaic} we  compare the MSF and NSF models to the reference $\Lambda$CDM model via the differences $\Delta$AIC and  $\Delta$BIC, whose values are qualitatively interpreted as described in \cite[][tables VI and VII]{Camarena:2018nbr} where more details can be found  \citep[see also][]{Basilakos:2019zsf}.
The results of this analysis  show that there is a little support for the MSF and NSF models (AIC) and there is strong evidence against the MSF and NSF models (BIC)---the differences are all positive. For completeness we also report the reduced $\chi^2$, which assesses the overall quality of the fit.

	\subsection{Common parameters and tensions}
	
	In figure~\ref{figc2} we show the $1\sigma$ and $2\sigma$
	confidence levels for the parameters that are in common with the $\Lambda$CDM model.
	Interestingly, the contours do not just widen but also shift towards higher values of $H_0$  and smaller values of $\Omega_m$, and this has the potential to solve the tension between high- and low-redshift observables as far as these parameters are concerned. Consequently, the MSF and NSF models may become observationally  interesting if the above mentioned tensions become more severe.

\subsection{Results with the local determination of $H_0$} \label{riessization}

In light of the tension between local and global determinations of the Hubble constant (see Section~\ref{hub}), it is interesting to repeat the previous analysis including the local determination of $H_0$ by \citet{Riess:2019cxk}, given in equation~\eqref{h01}.
The result is given in figure~\ref{figc22} and table~\ref{tab:res2} (the results regarding $S_8$ are explained in Section~\ref{sec:S8}).

As expected, the allowed values for $H_0$ become larger. In particular, the MSF and NSF models have an $H_0$ close to $70 \textrm{ km s}^{-1} \textrm{Mpc}^{-1}$, in better agreement with the result of equation~\eqref{h01}.
 Notice that due to the stronger statistical weight of the CMB prior as compared to equation~\eqref{h01}, the best-fit value of $H_0$ remains closer to the Planck inferred value.
This is the case also for our results relative to the concordance $\Lambda$CDM model. This is in agreement with \citet[][Planck Legacy Archive]{Aghanim:2018eyx}.
 Using the best-fit values of the cosmological parameters in the lower half of table~\ref{tab:res2}, we obtain the present value of equation of state 
	$w_{ \Phi}= -0.993^{+0.005} _{-0.005}$  and  $w_{\Phi}=-0.994^{+0.0035}_{-0.006}$ for MSF and NSF, respectively.
	We also obtain $F = 0.999^{+0.00004}_{-0.00004}$ at the present time for the NSF model. The results of model selection are given in table~\ref{tabaic2}. 
 Now, according to AIC, the results of background observables  indicate substantial support for MSF model and considerably less support for NSF model. In this context, the results of background and perturbation observables represent considerably less support for $\Lambda$CDM cosmology compare to MSF and NSF models. Note that the background observables used here include the local $H_0$ data. According to BIC and based on the background observables, there is a positive evidence against the MSF and strong evidence against the NSF model. While based on the combination of background and perturbation data, the BIC metric shows positive evidence against the NSF model.

Another interesting point that we see in the bottom panel of figure \ref{figc22} is that the inclusion of RSD data breaks degeneracies and allows for a tighter constrain on the scalar potential scale $\bar{k}$ and steepness parameter $\alpha$. To make this clearer, in figure \ref{kbar} we plot the quantity $f(z)\sigma_8(z)$ with $\bar{k}$ and $\alpha$ changed according to the background constraints ($\pm 3 \sigma$).
As expected, the quantity  $f(z)\sigma_8(z)$ is sensitive to the value of the parameters~$\bar{k}$ and $\alpha$.
 
 \subsection{Parameter combination $S_8$} \label{sec:S8}
	The combination of the total matter density $\Omega_{m0}$ and the clustering amplitude $\sigma_{80}$ as $S_8=\sigma_{80}\sqrt{\Omega_{m0}/0.3}$ is a useful quantity in order to compare the results of the cosmological model under study with observations. For example, in \cite{Kohlinger:2017sxk} the authors used 450 square degrees of the Kilo-Degree Survey (KiDS) imaging data to measure the weak gravitational lensing shear power spectrum with a quadratic estimator in two and three redshift bins. Note that the cosmological parameter that is constrained by the cosmic shear power spectrum data is $S_8$. The result reported by \cite{Kohlinger:2017sxk} is $S_8=0.651^{+0.058}_{-0.058}$, confirming the 3.2$\sigma$ tension  with constraints from the Planck CMB probe \citep{Ade:2015xua}. Here, in our analysis, we calculate the best-fit value as well as the confidence regions of the cosmological parameter $S_8$ using the combination of the background data (including local $H_0$) and  perturbation data. The results are presented in the last column of the lower half of Table \ref{tab:res2} and also in Figure \ref{fig:S8}. We observe  $1.8\sigma$, $2.0\sigma$ and $2.8\sigma$ deviations from the Planck constraint \citep{Aghanim:2018eyx} for MSF, NSF and $\Lambda$CDM cosmologies, respectively. We see that in the case of the MSF and NSF models the tension is alleviated as compared to the concordance $\Lambda$CDM model.
 
\section{CONCLUSIONS}\label{sec5}
In this paper we studied the minimally and non-minimally coupled scalar field models as  possible candidates for DE---the so-called scalar-tensor cosmology.
 After discussing the relevant equations, we showed the evolution of important cosmological quantities such as the scalar field equation of state and energy density, the Hubble rate and the effective gravitational constant.

We then confronted the two models with the latest supernova, BAO, CMB, $H(z)$ and RSD data and obtained updated constraints on their parameters.
We performed model selection using the basic information criteria (AIC and BIC).
We found that the $\Lambda$CDM model is strongly favored when the local determination of the Hubble constant is not considered and that this statement is weakened once local $H_0$ is included in the analysis. We computed the parameter combination $S_8=\sigma_{80}\sqrt{\Omega_{m0}/0.3}$ and observed the decrement of the tension with respect to Planck constraints in the case of minimally and non-minimally coupled scalar field models.
Finally, it is worth stressing that from cosmological data we obtained for the coupling constant between DE and gravity,  $\xi\simeq -0.06^{+0.19}_{-0.19}$ at the $1\sigma$ uncertainty. This constraints is approaching the one from solar system tests $|\xi| \lesssim 10^{-2}$  and comparable to the conformal value $\xi=1/6$ at the $1\sigma$ uncertainty.

	\section{Acknowledgments}
	The authors gratefully thank  Prof.~A.~A.~Sen for the valuable comments and recommendations which definitely helped to improve the readability and quality of the paper. It is also a pleasure to thank Ignacy Sawicki and Sohrab Rahvar for useful discussions. The work of ZD has been supported financially by the Iran Science Elites Federation. The work of VM has been partially supported by CNPq and FAPES.

\newpage	
	
	\bibliographystyle{mnras}
	\bibliography{ref}
	
	\appendix
	\section{Datasets} \label{datasets}
	\subsection{Hubble data} \label{hub}

Observations of supernovas Ia calibrated with Cepheid distances to SN Ia host galaxies provide the value of the Hubble constant \citep{Riess:2019cxk}:%
\footnote{See \citet{Camarena:2019moy} for a cosmology-independent re-determination of $H_0$ using the calibration by \citet{Riess:2019cxk}.}

	\begin{equation} \label{h01}
H_0 = 74.03 \pm 1.42 \textrm{ km s}^{-1} \textrm{Mpc}^{-1}
\end{equation}

On the other hand, the most recent analysis of the CMB temperature fluctuations constrains the current expansion rate to \citep{Aghanim:2018eyx}:
\begin{equation} \label{h02}
H_0 = 67.66 \pm 0.42 \textrm{ km s}^{-1} \textrm{Mpc}^{-1}
\end{equation}
These determinations are in a tension at \vale{$4.3\sigma$} and can strongly impact  model selection \citep{Camarena:2018nbr}.

Besides determining the present-day Hubble rate, it is possible to determine the Hubble rate at various redshifts (observational Hubble  data, OHD) using the galaxy differential age and radial BAO methods. 
	In our analysis, we use the former and, in particular, the 31 data points in the redshift range $0.07\leq z \leq 1.965$ given in \citep[][table I]{Marra:2017pst}. These data are uncorrelated with the BAO data points discussed below.
$\chi^2_h$ is written as:
	\begin{equation}
	\chi^2_h ({\bf p})=\sum_{i}\frac{\left[ H(z_i,{\bf p})-H_i \right]^2}{\sigma_i^2} \,.
	\end{equation}
%
We will consider the measurement of \eqref{h01} in Section~\ref{riessization}.

\subsection{Supernovas Type Ia}
Type Ia supernovas are a mature and powerful probe of the background dynamics thanks to their standardizable nature.
Here we use the binned Pantheon SNIa dataset~\citep{Scolnic:2017caz}, which contains 40 data points from $z = 0.014$ to $z = 1.61$.
We marginalize analytically over the absolute magnitude $M$; for details see  \citet[][Section IVD]{Camarena:2018nbr}.

	\subsection{Baryon acoustic oscillations}
	We also consider BAO data. The BAO peak has been detected very strongly and is an important background probe as the (linear) physics behind BAO is well understood. We use data from seven different surveys: 6dFGS, SDSS-LRG, BOSS-MGS, BOSS-LOWZ, WiggleZ, BOSS-CMASS, BOSS-DR12. See  \citet[][Section IVC]{Camarena:2018nbr} for details.

	\subsection{CMB data}
	The position of the CMB acoustic peak is useful
	to constrain DE models. The position of this peak is given
	by $(l_a , R, z_{\star} )$, where the shift parameter $R$ is defined as
	\begin{equation}
	R=\sqrt{\Omega_m}H_0 D_A(z_\star) \,,
	\end{equation}
	and the angular scale $l_a$ of the sound horizon at the time of decoupling is given by
	\begin{equation}
	l_a=\frac{\pi}{\Theta_a} \,,
	\end{equation}
	where $\Theta_a$, the angular scale of the first peak of the angular power spectrum of CMB anisotropies, is 
	\begin{equation}
	\Theta_a=\left[\int_{z_\star}^{\infty}c_s(z)\frac{dz}{H(z)}\right]\left[\int_{0}^{z_\star}\frac{dz}{H(z)}\right]^{-1} \,,
	\end{equation}
	where $z_{\star}$ refers to the redshift at the time of the last scattering.  For $z_{\star}$ we used \citep{Hu:1995en}:
	\begin{eqnarray}\label{eq:4.47}
	&&z_* = 1048(1+ 0.00124(\Omega_{b_0}h^2)^{-0.738})(1+ g_1(\Omega_{b_0}h^2)^{g_2}),\nonumber\\
	&&g_1 = 0.0783(\Omega_{b_0}h^2)^{-0.238}(1+ 39.5(\Omega_{b_0}h^2)^{0.763})^{-1},\nonumber\\
	&&g_2 = 0.560(1+ 21.1(\Omega_{b_0}h^2)^{1.81})^{-1}.
	\end{eqnarray}
		The $\chi^2_{\rm cmb}$ is given by
		\begin{equation}\label{eq:3.48}
		\chi^2_{\rm cmb}=\textbf{X}^T_{CMB} \textbf{C}_{CMB}^{-1} \textbf{X}_{CMB}.
		\end{equation}
		We used a prior on $(R, l_a, \Omega_{b_0}h^2)$  for the $w$CDM model derived from the Planck 2018 results \citep{Chen:2018dbv}:
		\[{\bf X}_{CMB}= \left(
		\begin{array}{l l l}
		R - 1.7493\\
		l_a - 301.462\\
		\Omega_{b_0}h^2 - 0.02239\\
		\end{array} \right) \]
		and
		\begin{equation}\label{eq:3.47}
		{\bf C}_{CMB}^{-1} = 
		\left({\begin{array}{ccc} 95242.51&-1367.03&1651932.7\\-1367.03&162.3696&5086.66\\1651932.7&5086.66&78905711.17\end{array}}\right).
		\end{equation}

	\subsection{Big Bang Nucleosynthesis}
	
	Big Bang Nucleosynthesis (BBN) provides a data point which constrains the physical baryon density $(\Omega_{b0} h^2)$ \citep{Mehrabi:2015kta}.
	The $\chi^2_{\rm bbn}$ is given by
	\begin{equation}
	\chi^2_{\rm bbn}=\frac{(\Omega_{b0}h^2-0.022)^2}{0.002^2} \,.
	\end{equation}
	\subsection{Redshift Space Distortions }
	Redshift space distortions (RSD) in galaxy redshift surveys measure the peculiar velocities of matter and thus infer the growth rate of cosmological perturbations on a
	range of redshifts and scales. RSD data have been useful to constrain the history of structure formation since about 2006, and will be crucial to understand the nature of DE using upcoming surveys \citep{Marra:2019lyc}.

	 RSD data allow us to constraint the combination $f\sigma_8(z)$~\citep{Song:2008qt} and consequently the cosmic growth index $\gamma$.
	 Here, we use the robust $f\sigma_8 (z)$ measurements from the ``Gold-2017'' compilation given in \citet[][table III]{Nesseris:2017vor}.
	 
In order to define $\chi^2_{\rm fs}$, first we define the following vector:
	\begin{equation}
	V^i(z_i,{\bf p})= f\sigma_{8,i} - q(z_i,\Omega_{dm} ^{fid}) f\sigma_8(z_i,{\bf p})  
	\label{VRSD} 
	\end{equation}
	where $ q(z,\Omega_{dm} ^{fid})=H(z)d_z(z)/H(z)^{fid}d_z(z)^{fid}$ is the fiducial correction factor for model dependence, so that $\chi^2_{\rm fs}$ can be written as
	\begin{equation}
	\chi^2_{\rm fs}=V^i \Sigma_{ij}^{-1}V^j \,.
	\label{chi2RSD1} 
	\end{equation}
	We remind that $f\sigma_8(z)=f(z) \, \sigma_8(z) $  is independent of galaxy
	bias and that $\sigma_{80}$ is the root-mean-square mass fluctuation in spheres with radius $8h^{-1}$ Mpc at~$z=0$.

			\label{lastpage}
\end{document}